\documentclass[preprint2]{aastex}
\usepackage{amsmath}
\pdfoutput=1
\begin{document}
\title{Saving super-Earths: Interplay between pebble accretion and type I migration}
\author{R.~Brasser\altaffilmark{1}, B.~Bitsch\altaffilmark{2} and S.~Matsumura\altaffilmark{3}}
\altaffiltext{1}{Earth Life Science Institute, Tokyo Institute of Technology, Meguro, Tokyo 152-8551, Japan}
\altaffiltext{2}{Department of Astronomy and Theoretical Physics, Lund University, Lund 22100, Sweden} 
\altaffiltext{3}{School of Science and Engineering, Division of Physics, Fulton Building, University of Dundee, Dundee 
DD1 4HN, UK}
\begin{abstract}
Overcoming type I migration and preventing low-mass planets from spiralling into the central star is a long-studied topic. It is well 
known that outward migration is possible in viscous-heated discs relatively close to the central star because the entropy gradient 
can be sufficiently steep that the positive corotation torque overcomes the negative Lindblad torque. Yet efficiently trapping planets 
in this region remains elusive. Here we study disc conditions that yield outward migration for low-mass planets under specific planet 
migration prescriptions. Using a steady-state disc model with a constant $\alpha$-viscosity, outward migration is only possible when 
the negative temperature gradient exceeds $\sim$0.87. We derive an implicit relation for the maximum mass at which outward migration 
is possible as a function of viscosity and disc scale height. We apply these criteria, using a simple power-law disc model, to 
planets that have reached their pebble isolation mass after an episode of rapid accretion. It is possible to trap planets with the 
pebble isolation mass farther than the inner edge of the disc provided that $\alpha_{\rm crit}\gtrsim 0.004$ for discs older than 
1~Myr. In very young discs the high temperature causes the planets to grow to masses exceeding the maximum for outward migration. 
As the disc evolves, these more massive planets often reach the central star, generally only towards the end of the disc's lifetime. 
Saving super-Earths is therefore a delicate interplay between disc viscosity, the opacity profile and the temperature gradient in the 
viscously heated inner disc.
\end{abstract}

\section{Introduction}
The topic of planet formation is one that has vexed modellers for many decades because it encompasses a range of physical processes 
that are tightly interlinked yet occur at completely different { length and time} scales. Planet formation takes place in 
protoplanetary discs that surround young newborn stars. Gas giants have to form within the lifetime of the protoplanetary disc, 
which is to say within a few million years \citep{Mam09}, while terrestrial planets, especially in our own solar system, take an order 
of magnitude longer until they are fully assembled. { For example, hafnium-tungsten dating indicates the Earth's core formed in 
about 10~Myr \citep{Yin2002}, and the terrestrial planets themselves in less than 100~Myr, consistent with early dynamical models 
\citep{C01}}. On the other hand, the time scale and formation mechanisms for super-Earth planets, i.e. planets that are a few times 
more massive than the Earth but whose bulk composition is still predominantly composed of rock { and possibly ice}, are not well 
understood. { Progress has been hampered by inadequate numerical approximations to planet migration, gas accretion and tidal 
damping forces from the gas disc. The long time scale, short orbital period and high number of bodies make N-body simulations 
extremely challenging and time-consuming} \citep{CN14,C14}. { Efforts to combine N-body simulations and chemical composition 
variation have been applied with moderate success to the terrestrial planets of the Solar System \citep{Bond10,Mat16}, but 
applications to exoplanets have so far been a more formidable task \citep{FC14,Mor14}}.\\

In the core accretion scenario of giant planets, a planetary core of about 10 Earth masses forms first and subsequently accretes an 
envelope of gas as soon as accretion of leftover solids is very low \citep{Pol96}. However, forming a planetary core of a few Earth 
masses through planetesimal accretion at large orbital distances in a disc with solar metallicity is not possible during the disc's 
lifetime. The planetary cores tend to scatter planetesimals in their vicinity rather than accrete them \citep{TI99}. In addition, 
planetesimal-driven migration will isolate some cores and prevent further accretion \citep{LD10}. However, recent models that consider 
the accretion of millimetre to centimetre-sized objects -- dubbed pebbles -- have solved the time scale problem for the core formation 
of giant planets \citep{Ormel10,LJ12,LD15a}. These works demonstrated that efficient core growth at large orbital distances is 
possible well within the lifetime of the protoplanetary discs.\\

The reason for the increased accretion rate of pebble accretion versus planetesimal accretion is the effect of gas drag. As 
pebbles lose their angular momentum due to friction with the sub-Keplerian gas, they drift sunwards. The rapid time scale for 
pebble migration and the strong headwinds they endure results in the pebble accretion cross-section of a planetary embryo being 
roughly equal to the radius of its Hill sphere rather than the gravitational cross-sectional radius. { However, this rapid 
accretion only works when the time it takes for the pebble to drift past the planet's orbit is longer than the stopping time 
\citep{Ormel10}. Otherwise, the pebbles just drift past the planet relatively untouched}. Planetesimals, on the other 
hand, are { generally scattered rather than accreted} because { the drag force that binds pebbles to the planet and results 
in accretion is virtually absent for typical-sized planetesimals}.\\

The growth of a planetary core via pebbles crucially depends on the pebble's size { in a complicated manner \citep{ida16}}. The 
typical pebble size is determined by an interplay between pebble growth and drift. Pebble accretion ends as soon as the planet starts 
to carve a partial annulus in the gas of the protoplanetary disc \citep{Lam14}, which changes the pressure gradient outside of the 
planetary orbit and thereby opens a corresponding annulus in the dust distribution of the disc \citep{PM6}. This ends the supply of 
pebbles to the planet. This is the point at which the planet has reached its pebble isolation mass, which strongly depends on the 
disc's aspect ratio. When the planet finishes accreting pebbles, it can contract a gaseous envelope, which allows the transition 
into runaway gas accretion as soon as the planetary envelope becomes more massive than the planetary core. The contraction time scale 
of the envelope is highly dependent on the mass of the planetary core, with a heavier core accelerating contraction \citep{I00}.\\

During their growth, the planets also migrate through the protoplanetary disc. Small planets that barely influence the structure of 
the disc undergo type I migration, while gap-opening planets undergo type II migration \citep{GT80,LP86,W97}. Type I migration is a 
combination of the Lindblad torque -- caused by the spiral waves generated in the disc by the planet -- and of the barotropic and 
entropy related corotation torques. The Lindblad torque is generally negative and thus causes starward (inward) migration, while 
the corotiaton torque can be positive and therefore cause outward migration if it is strong enough to overcome the negative Lindblad 
torque.\\

{ The corotation torque arises from material orbiting on horseshoe-orbits, which are co-rotating with the planet. Material 
circulating on orbits interior to the planet orbit the star slightly faster and thus catch up to the planet. Once they meet the 
material recedes from the planet again, due to the planet deflecting it to exterior orbits. If the disc interior to the planet is 
hotter than exterior, the material cools and expands adiabatically when being transported to outer orbits, generating a 
density difference across the planetary orbit. Similarly, material circulating on exterior orbits travels slower than the planet and 
are deflected inwards, where the gas heats up and adiabatically compresses. The density gradient causes the corotation torque on 
the planet, which can drive outward migration.\\

However, this torque is prone to saturation, where the effects described above can saturate and outward migration stops. Two 
effects are important here: (i) heating and cooling and (ii) viscosity. Without heating and cooling effects, the material orbiting on 
the horseshoe orbits would become fully mixed and the temperature gradient would disappear, resulting in the destruction of the over- 
and under-density in front of and behind the planet. The corotation torque thus saturates and outward migration is no longer possible. 
Heating and cooling is thus necessary to keep the temperature gradient in the corotation region, where a minimum temperature gradient 
is required to give this effect enough power to overcome the Lindblad torque and drive outward migration.\\

Viscosity, on the other hand, is needed to replenish the material inside the corotation region because this is where the planet 
extracts angular momentum from this region and the angular momentum of the corotation region is finite. An influx of new material 
inside the corotation region is thus needed to sustain outward migration. In order to supply the planet with more angular momentum at 
each U-turn the viscous timescale needed to replenish the corotation region must be shorter than the horseshoe libration timescale - 
which is the time taken for material to flow through the entire corotation region - because viscous replenishment must be faster than 
the material can complete a full cycle. Additionally, the viscous timescale must be longer then the U-turn time scale - which is the 
time the material needs to make a U-turn close to the planet - because a shorter viscous time scale would replace the material 
before it finishes its U-turn and prevent it from transferring angular momentum to the planet.\\

Note here that the viscous time scale depends on the width of the horseshoe region\footnote{The width of the horseshoe region scales 
with the square root of the planet's mass} squared, so it scales linearly with the planet's mass, while the libration and U-turn 
time scales are proportional to the inverse of the planet's mass \citep{BM08}. This indicates that the torques on more massive 
planets saturate, unless the viscosity is high.}\\

The strength of the torques also essentially depends on the radial gradients of surface density, temperature and entropy 
\citep{P11}. Generally, a strong radial gradient in temperature (and thus entropy, because the two are related) is needed to sustain 
outward migration { because it will cause a steeper density gradient across the planet's orbit and thereby increase the power of 
the corotation torque}. The delicate interplay between the torques can lead to regions in the protoplanetary disc where the total 
torque is zero and planets do not migrate. Such outward migration regions typically occur when the relative disc scale height $h=H/r$ 
drops with radial distance \citep{B13}. The planets could then, in principle, stay at these locations until the disc dissipates, if 
they do not grow to masses larger than the maximum mass at which outward migration can be sustained. Clearly, both planet growth and 
planet migration are related to the disc structure, so that the latter is of fundamental importance in understanding the mechanisms 
behind planet formation.\\

In \citet{Bitsch15}, planet growth via pebble accretion was combined with recent models of planet migration to study the 
formation of planets in evolving protoplanetary discs with a prescribed evolution and viscosity. In that work the survival of 
super-Earth planets at orbital distances greater than 1~AU was only possible when the planets remained in the region of outward 
migration, but at the same time did not outgrow it \citep{C17}. Here we shall expand on this work by studying the interplay between 
planetary growth via pebble accretion and type I migration for super-Earths (i.e. low-mass planets) for a variety of disc parameters. 
The aim of this study is to explore a finite set of disc parameters and determine which set allows a suite of low-mass planets formed 
through pebble accretion to survive the migration instead of colliding with the central star.\\

Our work is structured as follows. In section 2, we review the foundations of type I migration and study the interplay between 
diffusion and viscosity on planet migration and growth. In section 3, we study the influence of different temperature gradients and 
viscosity on the migration and survival of super-Earth planets. We then engage in a short discussion in Section 4, and present our 
conclusions in Section 5 followed by an Appendix summarising pebble accretion.

\section{Theory of planet migration}
In this work we will focus on a simple, commonly-used model that relies on the $\alpha$ viscosity to be constant \citep{SS73}. More 
complex disc models involving MRI and/or disc winds \citep{Bai16,Suz16} would deviate from the single $\alpha$ approach used here and 
are reserved for future studies.\\

We assume a steady accretion rate of the disc gas onto the Sun. The gas accretion rate is related to the gas surface density and scale 
height via \citep{P81}
\begin{equation}
\dot{M}_*=3\pi \Sigma \nu = 3\pi \alpha \Sigma H^2 \Omega_{\rm K},
\label{eq:dotmstar}
\end{equation}
where $\Sigma$ is the gas surface density, $\nu$ is the radially-dependent viscosity, $H$ is the disc scale height and $\Omega_{\rm 
K}$ is the Kepler frequency. The $\alpha$-viscosity is assumed to be constant so that $\nu=\alpha c_s^2\Omega_{\rm K}^{-1}$ 
\citep{SS73}. The disc scale height is related to the temperature via $H=c_s/\Omega_{\rm K}$ where $c_s=(k_BT/\mu m_p)$ is the 
isothermal sound speed, $k_B$ is the Boltzmann constant and $\mu=2.3$ is the mean atomic mass of the gas.\\

\subsection{Basic equations}
For this study we investigate the migration of a low-mass planet according to the formulation of \citet{P11}. The normalised torque on 
the planet is generally a function of four parameters, but in the derivation below we can narrow this down to three. The normalised 
torque is given by
\begin{equation}
 \frac{\Gamma}{\Gamma_0} = \Gamma_{\rm c} + \Gamma_{\rm L} = \Gamma_{\rm c,baro} + \Gamma_{\rm c,ent} + \Gamma_{\rm L}
\end{equation}
where $\Gamma_{\rm c}$ and  $\Gamma_{\rm L}$ are the corotation and Lindblad torques respectively { and $\Gamma_0= 
(m_p/m_\odot)^2(H/r)^{-2}\Sigma\Omega_{\rm K}^2$ is a normalisation constant}. The corotation torque is further subdivided into 
contributions from the barotropic and entropy parts. The corotation torque in a non-isothermal disc with thermal diffusion becomes 
\begin{equation}
 \Gamma_{\rm c,baro} =  F(p_\nu)G(p_\nu)\Gamma_{\rm hs,baro} + [1-K(p_\nu)]\Gamma_{\rm c,lin,baro},
\end{equation}
for the barotropic part while the entropy contribution is
\begin{eqnarray}
 \Gamma_{\rm c,ent} &=&  F(p_\nu)F(p_\chi)[G(p_\nu)G(p_\chi)]^{1/2}\Gamma_{\rm hs,ent}  \\
 &+&\{[1-K(p_\nu)][1-K(p_\chi)]\}^{1/2}\Gamma_{\rm c,lin,ent}.\nonumber
 \label{eq:entropypart}
\end{eqnarray}
{ The 'lin' and 'hs' subscripts indicate the Lindblad and horseshoe (corotation) parts of the torque respectively (see below).} The 
functions $F(p)$, $G(p)$ and $K(p)$ are defined as \citep{P11}

\begin{eqnarray}
F(p) &=& \frac{8I_{4/3}(p)}{3pI_{1/3}(p)+\frac{9}{2}p^2I_{4/3}(p)} \approx \frac{1}{1+(p/1.3)^2}\nonumber \\
G(p) 
&=&\frac{16}{25}\Bigl(\frac{45\pi}{8}\Bigr)^{3/4}\Bigl[1-\Theta\Bigl(p-\sqrt{\frac{8}{ 45\pi } }\Bigr)\Bigr]p^{3/2}\\
&+&\Theta\Bigl(p-\sqrt{\frac{8}{45\pi}}\Bigr)\Bigl[1-\frac{9}{25}\Bigl(\frac{8}{45\pi}\Bigr)^{4/3}p^{-8/3}\Bigr],\nonumber \\
K(p) 
&=&\frac{16}{25}\Bigl(\frac{45\pi}{28}\Bigr)^{3/4}\Bigl[1-\Theta\Bigl(p-\sqrt{\frac{28}{ 45\pi } }\Bigr)\Bigr]p^{3/2}\nonumber \\
&+&\Theta\Bigl(p-\sqrt{\frac{28}{45\pi}}\Bigr)\Bigl[1-\frac{9}{25}\Bigl(\frac{28}{45\pi}\Bigr)^{4/3}p^{-8/3}\Bigr]. \nonumber
\label{FGK}
\end{eqnarray}
In the above equations, $I_\nu(p)$ are modified Bessel functions of the first kind and $\Theta(x)$ is the heaviside step function. The 
remaining torque terms are functions of the temperature and surface density gradients $q=-\frac{d\ln T}{d\ln r}$ and $s=-\frac{d\ln 
\Sigma}{d \ln r}$ so that $T(r)\propto r^{-q}$ and $\Sigma \propto r^{-s}$. In steady state, $q+s=\frac{3}{2}$ and only one 
is independent. It follows that $\frac{d\ln H}{d \ln r} = \frac{3}{2}-\frac{1}{2}q$. The remaining contributions to the 
torque are then given by \citep{P11}

\begin{eqnarray}
 \Gamma_{\rm L} &=& -2.5 -1.7q +0.1s = -2.35 -1.8q \nonumber \\
 \Gamma_{\rm hs,baro} &=& 1.1(3/2-s) = 1.1q \nonumber \\
 \Gamma_{\rm c,lin,baro} &=& 0.7(3/2-s) = 0.7q \nonumber \\
 \Gamma_{\rm hs,ent} &=& \frac{7.9\xi}{\gamma} = 5.6\Bigl(\frac{7}{5}q-\frac{3}{5}\Bigr)\nonumber \\
 \Gamma_{\rm c,lin,ent} &=& \Bigl(2.2-\frac{1.4}{\gamma}\Bigr)\xi = 0.8\Bigl(\frac{7}{5}q-\frac{3}{5}\Bigr),
 \label{eq:gammacomps}
\end{eqnarray}
where $\xi=-\frac{d\ln S}{d\ln r}=q-(\gamma-1)s=\frac{7}{5}q-\frac{3}{5}$ is the negative entropy gradient and $\gamma$ is the ratio 
of specific heats, which is set to $7/5$. When $q=3/7$, which occurs in the outer parts of the disc where the disc's temperature 
is dominated by stellar irradiation, the entropy gradient is zero, implying that outward migration is only possible in the inner parts 
of the disc dominated by viscous heating \citep{B13}.\\

\begin{figure}
\resizebox{\hsize}{!}{\includegraphics{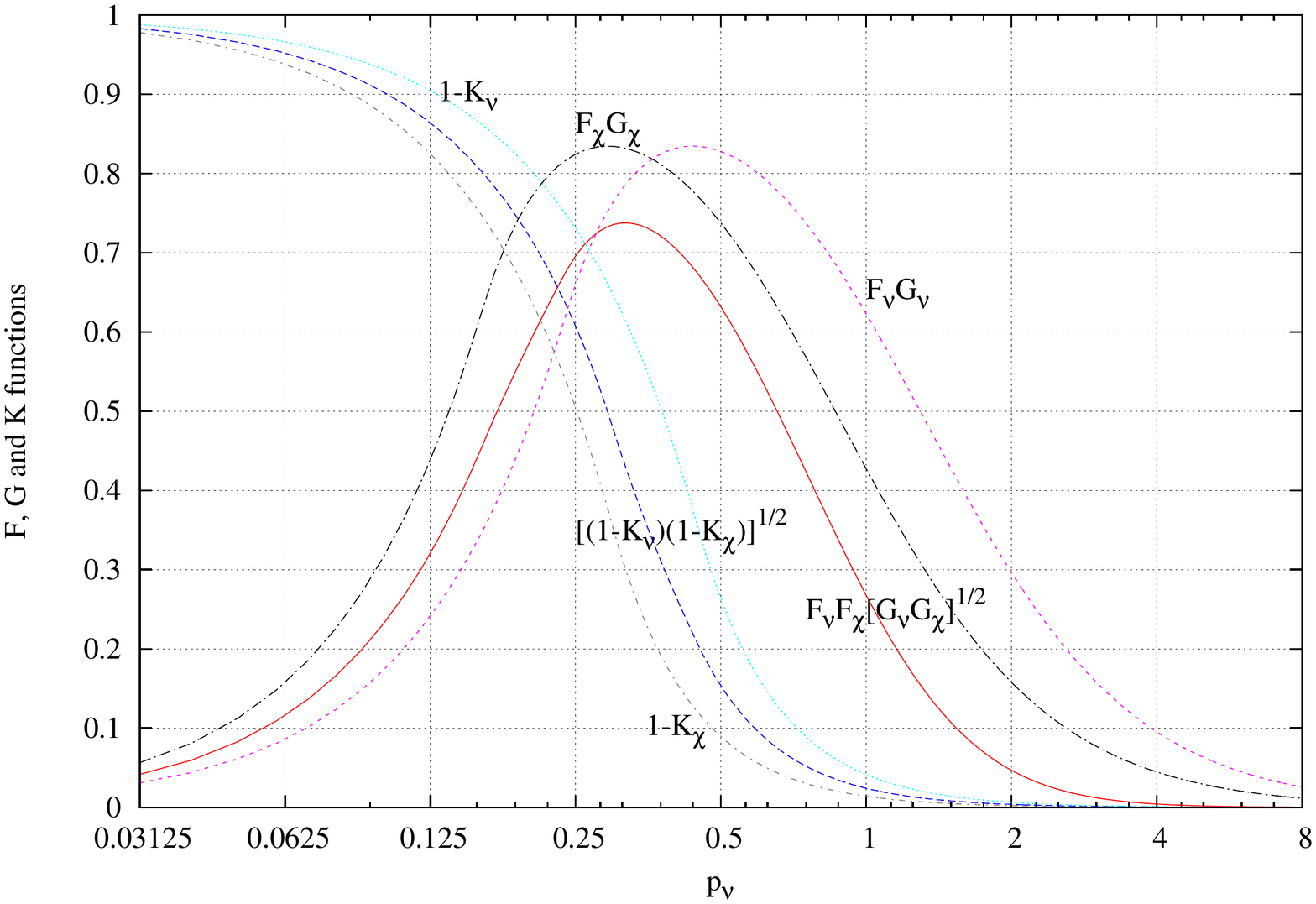}}
\caption{Graph of the various F, G and K functions as they appear in the formulae for the corotation torque. We set 
$\chi=\nu$. The torque is strongest when these functions are simultaneously at a global maximum, which occurs when $p_\nu \sim 0.25$.}
\label{fig:comps}
\end{figure}

The parameter $p_\nu$ is defined as \citep{P11}
\begin{equation}
p_\nu = \frac{1}{3}rx_{\rm s}\sqrt{\frac{2\Omega_Kx_{\rm s}}{\pi\nu}}
\label{eq:pnu}
\end{equation}
with $x_{\rm s}=(\frac{m_p}{m_*})^{1/2}h^{-1/2}$ being the horseshoe half-width. Ultimately

\begin{eqnarray}
\label{eq:pnumr}
p_\nu &=& \frac{2}{3}(2\pi\alpha)^{-1/2}\Bigl(\frac{m_p}{m_*}\Bigr)^{3/4}h^{-7/4} \\
 &=&\frac{2}{3}(2\pi\alpha)^{-1/2}\Bigl(\frac{m_p}{m_*}\Bigr)^{3/4}h_0^{-7/4}\Bigl(\frac{r}{1\,{\rm AU}}\Bigr)^{-7/8(1-q)} \nonumber
\end{eqnarray}
where $h=H/r$ is the reduced disc scale height and we used the fact that in steady state $\frac{d \ln h}{d\ln r} = \frac{1}{2}(1-q)$. 
The value $h_0$ is the disc scale height at 1~AU (or any other unit of distance). When $q=1$ $p_\nu$ has no dependence on distance.\\ 

The parameter $p_\chi$ in equation (\ref{FGK}) is given by $p_\chi= \frac{3}{2}(\frac{\nu}{\chi})^{1/2}p_\nu$ and $\chi$ is the 
coefficient of thermal diffusion. It is related to the temperature, surface density and opacity of the disc via

\begin{equation}
 \chi=\frac{16\gamma(\gamma-1)\sigma_{\rm SB}T^4}{3\kappa\rho_{\rm g}^2H^2\Omega_{\rm K}^2}=\frac{32\pi\gamma(\gamma-1)\sigma_{\rm 
SB}T^4}{3\kappa\Sigma^2\Omega_{\rm K}^2}.
\end{equation}
Here $\kappa$ is the Rosseland mean opacity, $\rho_{\rm g}=\frac{\Sigma}{\sqrt{2\pi}H}$ is the gas midplane density, and $\sigma_{\rm 
SB}$ is the Stefan-Boltzmann radiation constant and we inserted a missing factor 4 from \cite{P11}. Since $\chi$ and $\nu$ have the 
same dimensions, one may think of $\chi$ as a diffusive viscosity and then subsequently define a parameter $\chi=\alpha_\chi 
H^2\Omega_{\rm K}$, analogous to the $\alpha$ parameter $\nu=\alpha H^2 \Omega_{\rm K}$. Then $\frac{\nu}{\chi} = 
\frac{\alpha}{\alpha_\chi}$, which is mostly a function of the opacity. { For the sake of simplicity, in the rest of the paper we 
shall usually assume $\nu=\chi$, which is valid in optically { thick} regions, unless stated otherwise.}

\begin{figure}
\resizebox{\hsize}{!}{\includegraphics[]{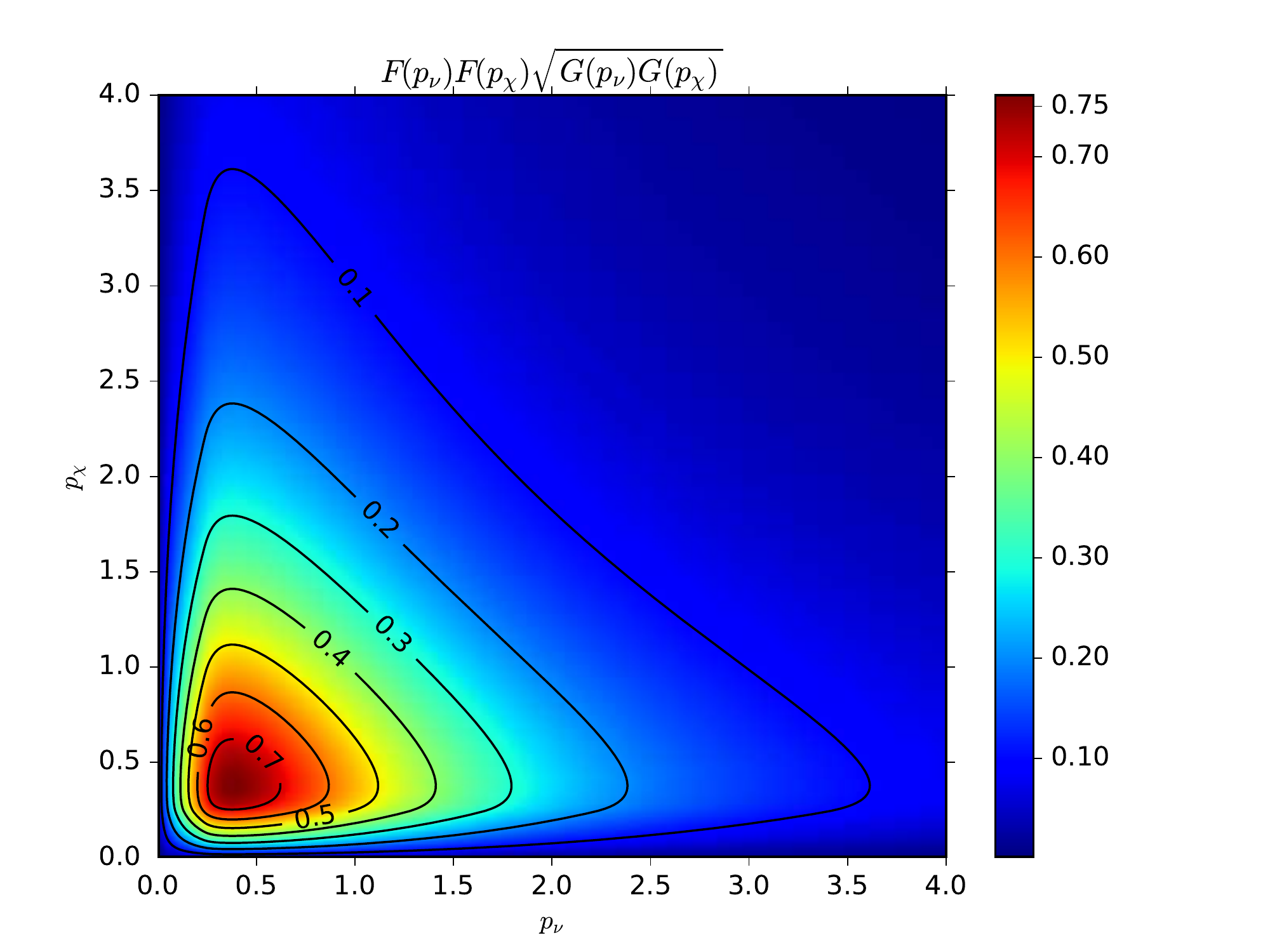}}
\caption{Contour image of $F(p_\nu)F(p_\chi)[G(p_\nu)G(p_\chi)]^{1/2}$ as a function of $p_\nu$ and $p_\chi$. The outward 
component of the torque is strongest when $F(p_\nu)F(p_\chi)[G(p_\nu)G(p_\chi)]^{1/2}$ is at its maximum, which occurs when 
$\chi=\frac{9}{4}\nu$ and at relatively small values of $p_\nu$ and $p_\chi$. The temperature gradient is $q=9/10$.}
\label{fig:comps2}
\end{figure}

\subsection{Conditions for outward migration}
At low eccentricity outward migration occurs when the positive corotation torque overcomes the negative Lindblad torque, and thus 
depends on the values of $q$, $p_\nu$ and $p_\chi$. However, since $p_\chi$ is a (non-fixed) multiple of $p_\nu$, we can analyse the 
above functions to find when the outward torque is a maximum as a function of $p_\nu$ and $q$ only if we fix $\frac{\nu}{\chi}$. In 
Fig.~\ref{fig:comps} we show the various combinations of F, G and K functions with $\nu=\chi$ as they appear in the corotation torque. 
The $1-K(p)$ functions are monotonically decreasing while the product $F(p)G(p)$ can often be approximated as a log-normal function. 
The product $F(p_\nu)G(p_\nu)$ and $F(p_\chi)G(p_\chi)$ have the same underlying distribution and maximum value when $p_\nu=p_\chi$, 
but the mode is shifted towards lower values of $p$ as $\frac{p_\chi}{p_\nu}$ increases. The torque is maximal when $p_\nu \sim 0.25$ 
because there the curves $F(p_\nu)F(p_\chi)[G(p_\nu)G(p_\chi)]^{1/2}$ and $\{[1-K(p_\nu)][1-K(p_\chi)]\}^{1/2}$ intersect, although 
the precise location of this intersection depends on the relative value of $\nu$ versus $\chi$.\\

{ The term $F(p_\nu)F(p_\chi)[G(p_\nu)G(p_\chi)]^{1/2}$ determines the saturation and the strength of the entropy part of the 
horseshoe torque (see equations~\ref{eq:entropypart}) and is mostly responsible for the outward migration of the planet.} The 
behaviour of the product $F(p_\nu)F(p_\chi)[G(p_\nu)G(p_\chi)]^{1/2}$ is shown in Fig.~\ref{fig:comps2} as a function of $p_\nu$ and 
$p_\chi$. The maximum occurs when $\chi=\frac{9}{4}\nu$. The product $\{[1-K(p_\nu)][1-K(p_\chi)]\}^{1/2}$ is monotonically 
decreasing in both $p_\nu$ and $p_\chi$ so that its global maximum is near the origin and thus is not displayed here. From 
Fig.~\ref{fig:comps2} one may infer that the outward torque is maximal when both $p_\nu$ and $p_\chi$ are in the range 0.3-0.7, 
setting limitations on the ratio $\frac{\nu}{\chi}$ and thus on the disc opacity { for which outward migration is feasible.}\\

\begin{figure}[ht]
\resizebox{\hsize}{!}{\includegraphics[]{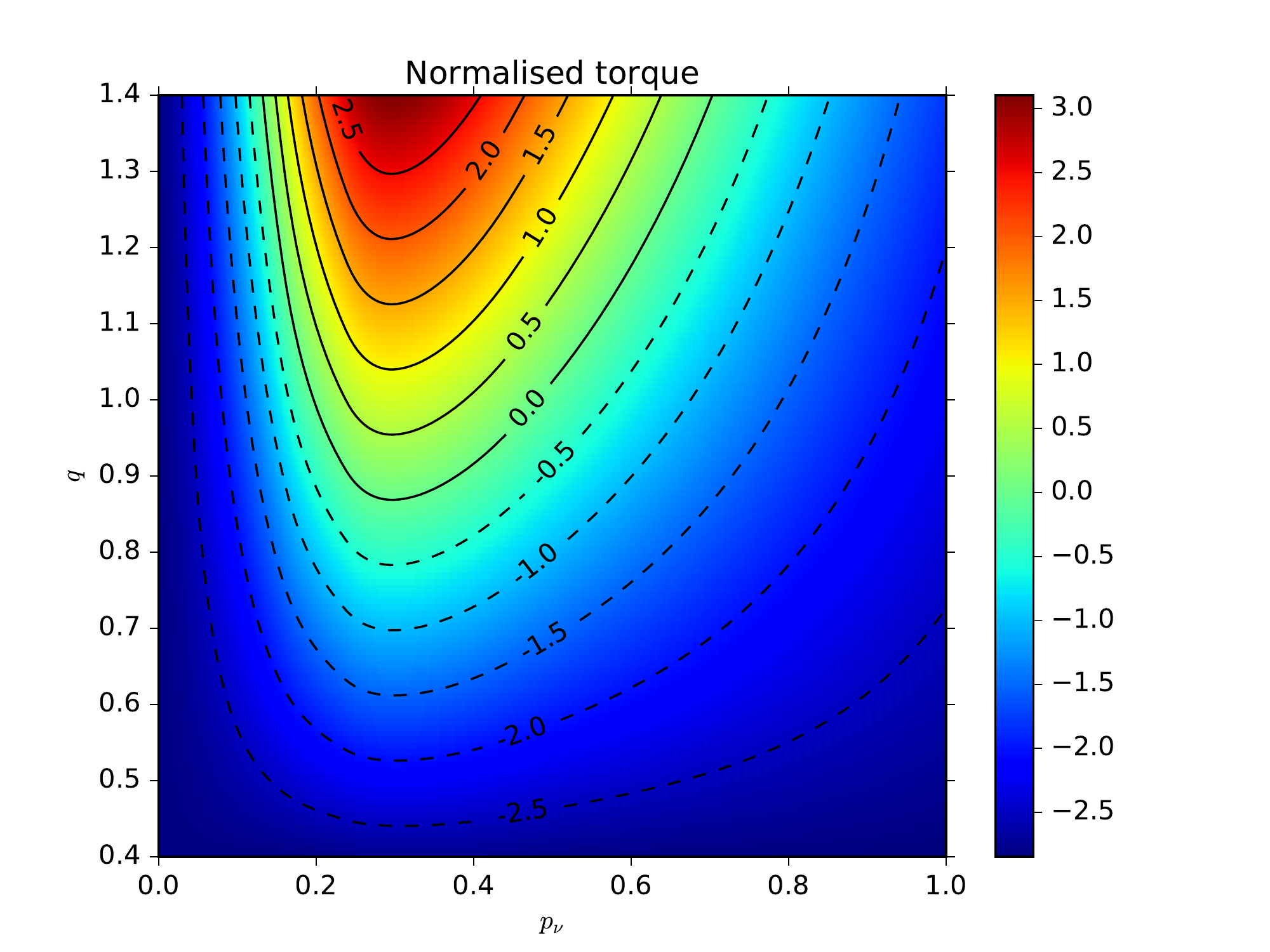}}
\caption{Contour image of the normalised torque $\Gamma$ as a function of the temperature gradient, $q$, and $p_\nu$ { (which 
itself depends on $q$; see equation~\ref{eq:pnumr})}. We have set $\chi=\nu$. The total torque is only positive in a narrow range of 
values of $p_\nu$ and only when $q\gtrsim 0.87$.}
\label{fig:pqtorque}
\end{figure}
The condition for outward migration does not only depend on $p_\nu$ and $p_\chi$, but also on the temperature gradient, $q$, through 
equations~\ref{eq:gammacomps}. A contour plot of the influence of the temperature gradient on the total torque is shown in 
Fig.~\ref{fig:pqtorque}, for $\chi=\nu$. Outward migration is only possible inside the contour of zero torque, and only occurs for a 
narrow range in $p_\nu$ and when $q \gtrsim 0.87$. \\

{ In the inner disc the temperature gradient is determined by viscous heating balanced with radiative cooling, and the 
cooling is inversely proportional to the opacity. The opacity gradient thus determines the cooling gradient and with it, the 
temperature structure of the disc. Outward migration is related not only to the temperature gradient (where a steeper slope favours 
outward migration), but also to the surface density gradient, where a shallow slope is preferred for outward migration. In a 
steady-state accretion disc, where the negative density and temperature gradients add up to 3/2, $q\gtrsim 0.87$ is needed to sustain 
outward migration, which is slightly lower than the typical temperature gradient of a viscous-heated and irradiated disc 
\citep{GL07,C09,B13}. In disc models with heating and cooling, $q$ can be as high as 1.2 in the viscous part through 
self-shadowing \citep{B15}, so we expect in the viscous region that $0.85 \lesssim q \lesssim 1.2$. The outer part of the disc is 
dominated by stellar heating, so that as a consequence of the heating/cooling balance with stellar irradiation $q=3/7$ always 
\citep{CG97}, though self-consistent models give a slightly steeper gradient of 4/7 in the outer parts of the shadowed region, that 
then become shallower, reach the 3/7 profile at larger orbital distances. \citep{B13}. Therefore, outward migration is unlikely to 
occur in the outer region of the disc.} \\

Increasing $\chi$ with respect to $\nu$ will shift the region of positive torque slightly towards lower $p_\nu$ and $q$, but once 
$\chi > \frac{9}{4}\nu$ the region of outward migration is confined to higher values of $q$ because the maximum of 
$F(p_\nu)F(p_\chi)[G(p_\nu)G(p_\chi)]^{1/2}$ are lower and the entropy contribution is stronger than the barotropic one (see 
equations~\ref{eq:gammacomps}). Thus, the region(s) of outward migration is obtained from solving $\Gamma=0$ for 
$p_\nu$ given $q$ and $p_\chi$, and then applying equation~(\ref{eq:pnumr}).\\

So far we have only parametrised the region of outward migration for a fixed ratio of $\frac{\nu}{\chi}$. However, the opacity of the 
disc is not a constant value as a function of the distance to the star, and thus we expect the above ratio to vary with distance, and 
time. Since the maximum outward torque occurs when $\chi=\frac{9}{4}\nu$, it is obvious that variations in the opacity do not 
necessarily help in increasing the planetary mass and distance range for which outward migration is possible, { though the same 
cannot be said for the opacity gradient}. We shall return to the role of opacity later. \\

In the next subsection we focus on the effect of pebble accretion and how to effectively prevent planets from spiralling all the way 
to the star.

\subsection{Outward migration at the pebble accretion isolation mass}
Apart from inside the snow line, where the pebbles may lose their volatiles quickly and they become coupled to the gas, pebble 
accretion is capable of producing planets of a few Earth masses in less than a million years \citep{Lam14,ida16}; generally 
much shorter than the lifetime of the gas disc. It is thought that pebble accretion ceases to be effective when the planet has reached 
what is termed the pebble isolation mass \citep{Lam14}. Due to angular momentum transfer between the planet and the disc, the material 
around the planetary orbit is pushed aside, which generates a pressure bump outside of the planetary orbit, trapping all pebbles. The 
pebbles can thus no longer reach the planet and pebble accretion self terminates. The pebble isolation mass is approximately given by 
\citep{Lam14}

\begin{equation}
 m_{\rm iso} \sim 20\Bigl(\frac{h}{0.05}\Bigr)^3\,M_\oplus.
\end{equation}
{ \citet{Lam14} briefly discuss whether the pebble isolation mass has an $\alpha$ dependence, and while they do not investigate 
this in great detail, they conclude that a weak dependence is possible. In what follows below, we assume no such dependence for the 
sake of simplicity, but we shall return to a possible weak dependence in Section~4.} \\

The dependence of the pebble isolation mass on the (reduced) scale height as $h^3$ is steeper than that of $p_\nu$, which goes as 
$h^{7/3}$ -- see equation~(\ref{eq:pnumr}). Since the total torque depends on $p_\nu$ (and $q$ and $p_\chi$), which itself is a 
function of both the planet mass and the disc scale height (and the viscosity), it is possible to determine where the contour of zero 
torque as a function of $h$ and $m_p$ intersects that of $m_{\rm iso}$ for a given value of $\alpha$ and $q$. Equating the isolation 
mass to the maximum mass for which the torque is zero and solved for $h$ using equation~(\ref{eq:pnumr}) results in
\begin{equation}
h_{\rm crit} = \frac{125\pi\sqrt{3}}{16}\alpha p_\nu^2,
\label{eq:hcrit}
\end{equation}
where the value of $p_\nu$ comes from Fig.~\ref{fig:pqtorque} { and we used the approximation that $m_\oplus/m_\odot = 3 \times 
10^{-6}$}. This is a powerful diagnostic from which one can quickly narrow down what regions of the disc support outward migration and 
what disc conditions are necessary to trap planets prevent them from reaching the central star. { For example, the typical maximum 
value of $p_\nu$ for which outward migration occurs is $p_\nu \sim 0.6$, which results in a maximum scale height of $h_{\rm crit,max} 
\sim 0.015$ when $\alpha=10^{-3}$. A corresponding minimum value occurs when $p_\nu \sim 0.15$ and $h_{\rm crit,min} \sim 0.001$ when 
$\alpha=10^{-3}$. A potential weak dependence of $m_{\rm iso}$ on $\alpha$ of the form $m_{\rm iso} \propto 
(\frac{\alpha}{10^{-3}})^\beta$ with $\beta \ll 1$ would lower $h_{\rm crit,max}$ when $\alpha>10^{-3}$ and increase it when 
$\alpha<10^{-3}$.}\\

\begin{figure*}
\centering
  \begin{tabular}{@{}ccc@{}}
   \includegraphics[width=.48\textwidth]{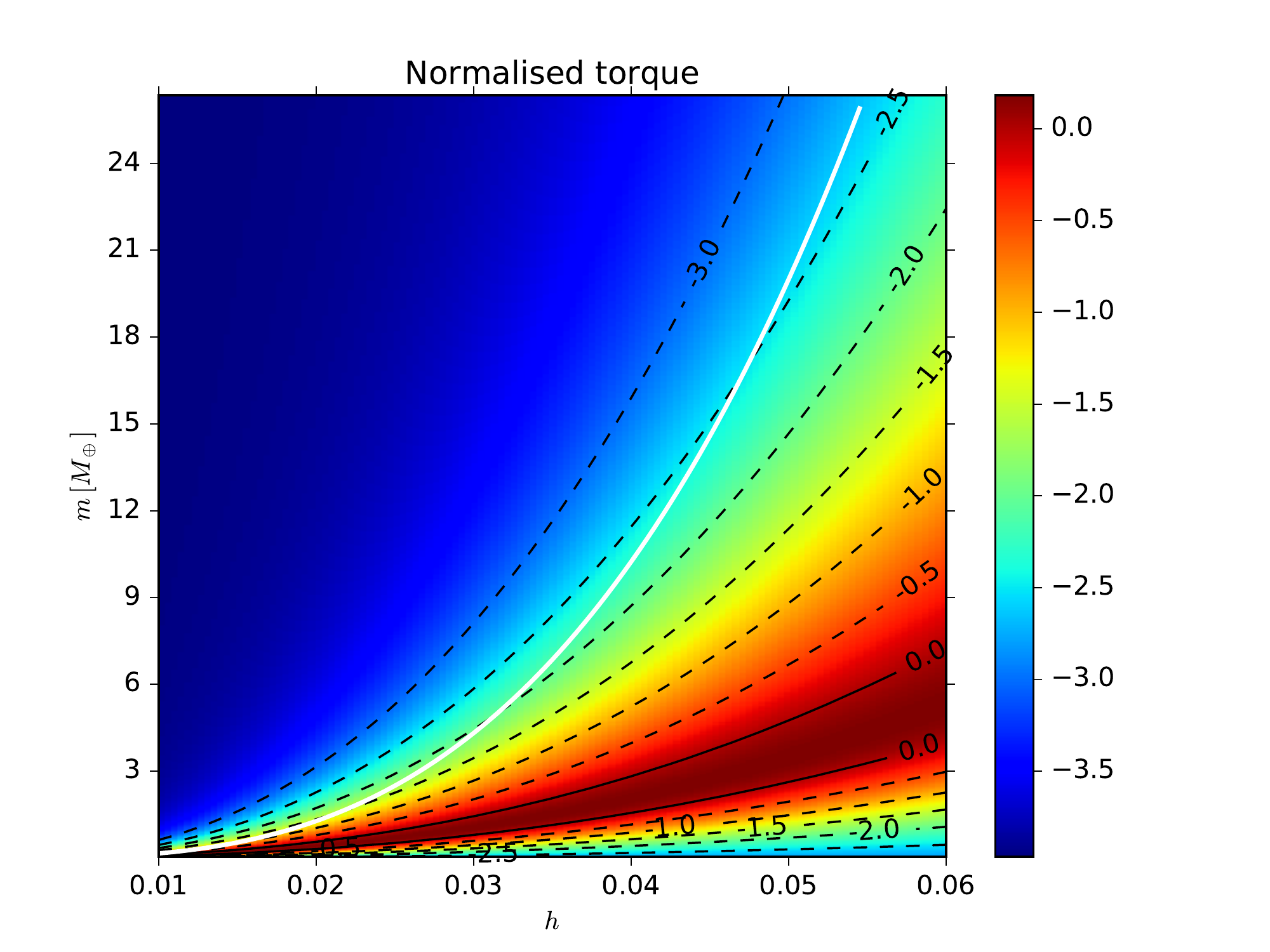} &
   \includegraphics[width=.48\textwidth]{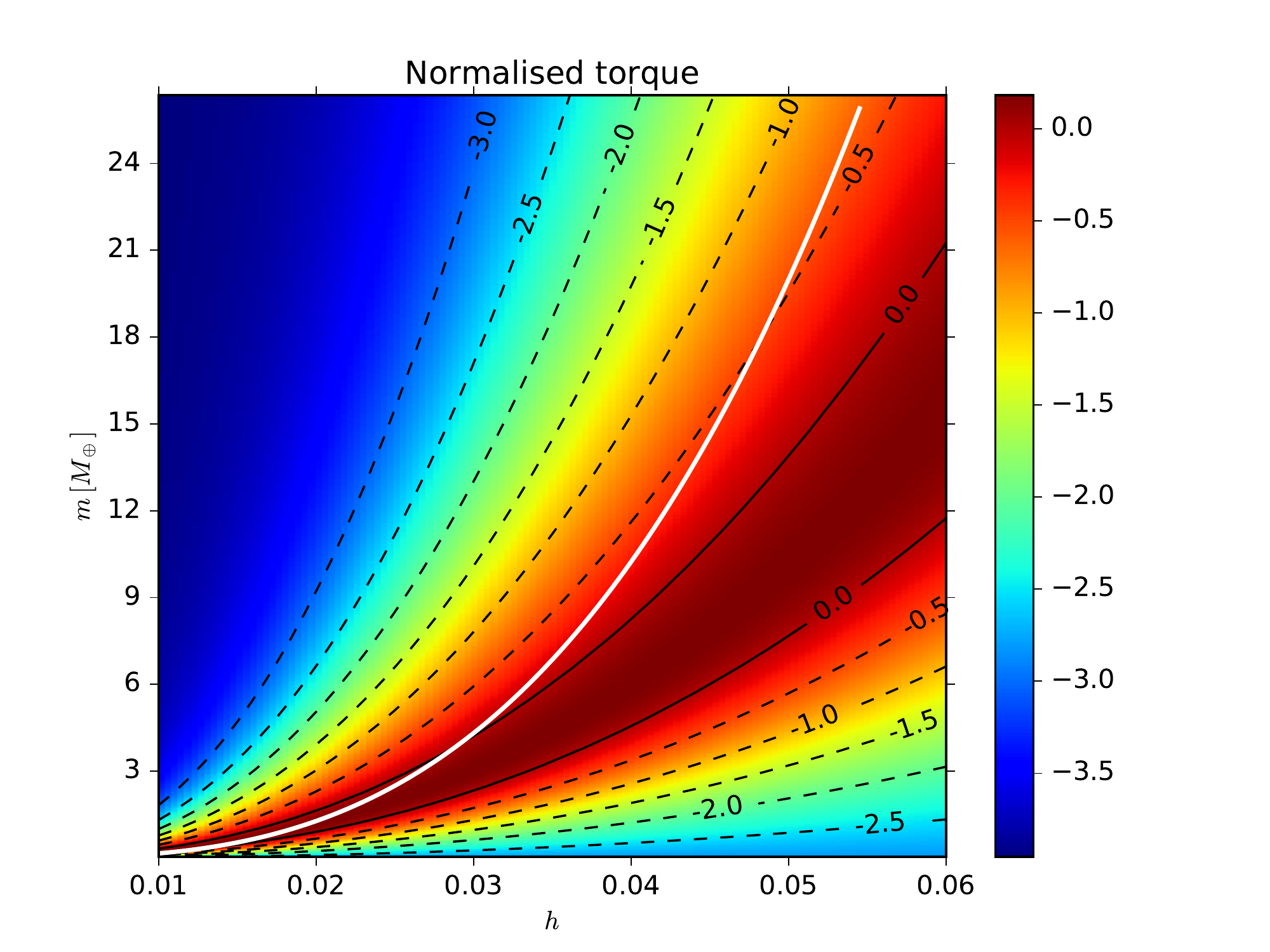} \\
   \includegraphics[width=.48\textwidth]{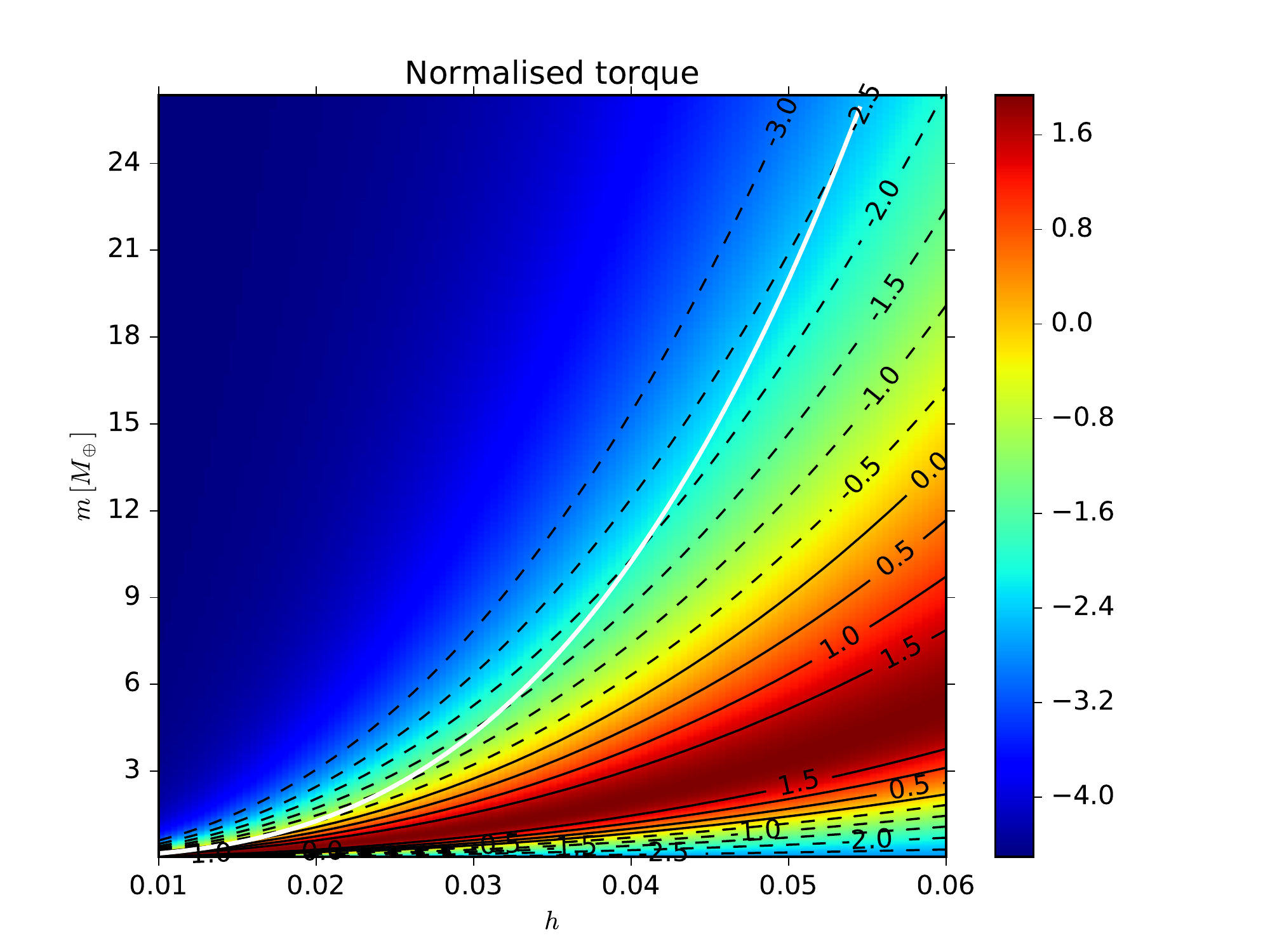} &
   \includegraphics[width=.48\textwidth]{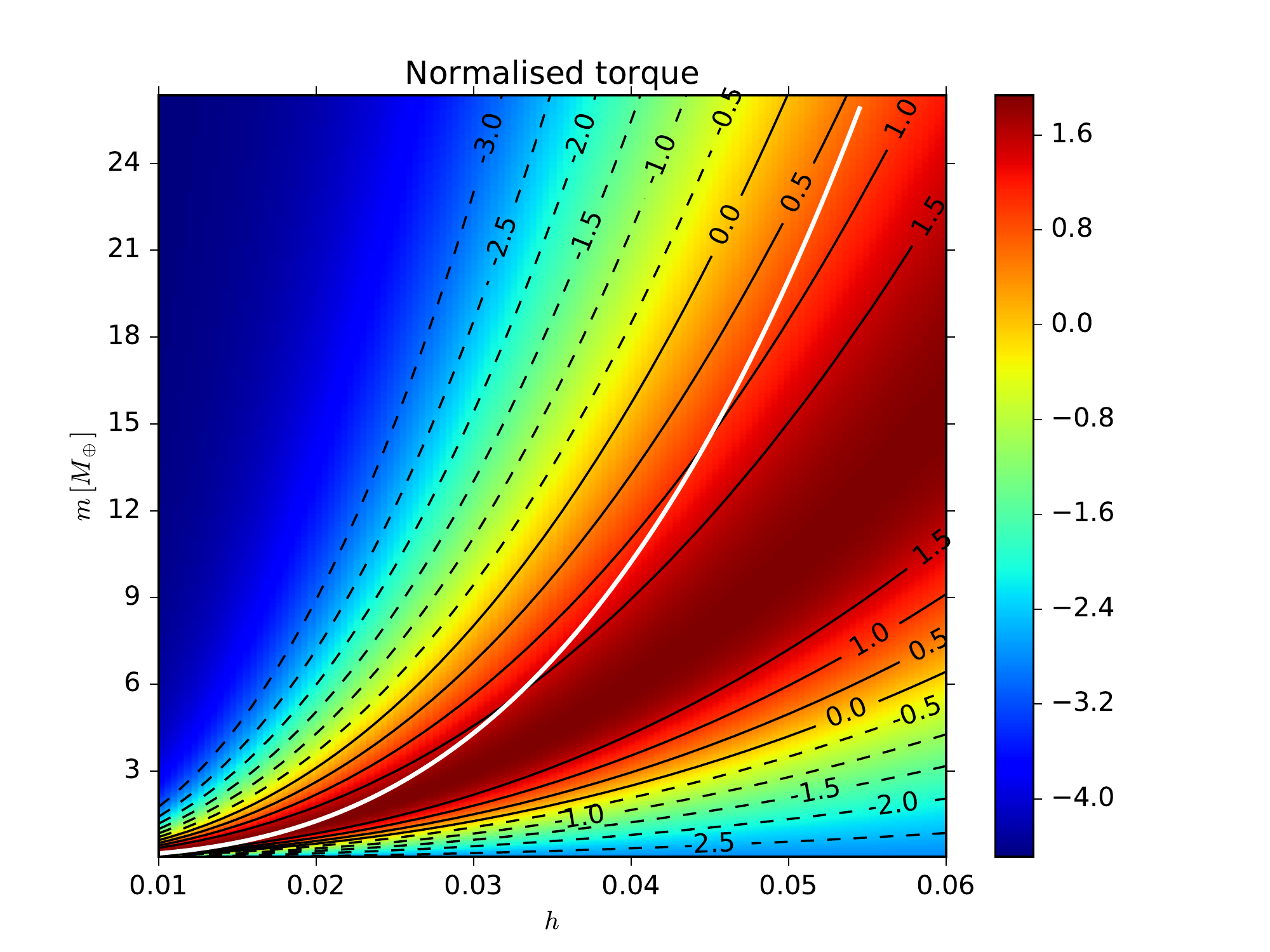}
  \end{tabular}
\caption{Contours of the total migration torque as a function of reduced disc scale height and planet mass for a given value of 
$\alpha$ and $q$. We have set $\chi=\nu$. In the left column $\alpha=10^{-3}$ and in the right column $\alpha=5\times10^{-3}$. The 
temperature gradient is 9/10 at the top row and 6/5 in the bottom row. The white line plots the planet isolation mass. Note that 
trapping planets at the isolation mass in the outward migration region is difficult with $\alpha=10^{-3}$ but becomes possible when 
$\alpha=5\times10^{-3}$.}
\label{fig:qrmsol}
\end{figure*}

We have plotted a number of cases in Fig.~\ref{fig:qrmsol} for various values of $\alpha$ and $q$. The left column has 
$\alpha=10^{-3}$ and the right column has $\alpha=5 \times 10^{-3}$. The top row has $q=9/10$ and $q=6/5$ in the bottom row. We set 
$\chi=\nu$. The white curve depicts the pebble isolation mass. \\

{ The growth and migration of a typical low-mass planet on these plots is similar to the following. With $\alpha$ fixed, planets 
initially grow at fixed $h$ and at some point in time begin to migrate inwards. When this happens $h$ decreases while $m_p$ still 
increases. Eventually the planet will reach the line where the torque is zero and gets trapped at the outer edge of the positive 
torque region. This entrapment could happen before or after it has reached its isolation mass. As time evolves the disc cools (see 
equation~(\ref{eq:T_visirr}) below), which implies that $h$ decreases. Meanwhile the planet's mass remains fixed or still increases. 
This combination of disc evolution and possible continued planet growth implies that eventually the planet evolves out of the positive 
torque zone and migrates towards the central star (unless $h$ or $\alpha$ somehow increases).\\}

It is clear from the plots on the left side that stalling planets in the outward region of migration when $\alpha=10^{-3}$ is only 
possible at very low isolation masses and scale heights ($h_{\rm crit, max} \lesssim 0.015$). On the other hand, when $\alpha=5 \times 
10^{-3}$, trapping planets in the outward migration region is possible even when $q=9/10$, provided the disc is thin ($h \lesssim 
0.03$). From the plot it is evident that the planet isolation mass at which outward migration occurs increases for increasing 
temperature gradient. This trend suggests there is a critical value of $\alpha$ where it is possible to prevent planets from reaching 
the central star for a given temperature gradient. We determine this threshold value of $\alpha$ and whether or not we can trap 
planets in the outward migration region in the next section.

\section{Results}
In this section we present the results of numerical simulations of single planet formation with pebble accretion as well as what disc 
conditions are needed to trap planets in the outward migration region.\\

\subsection{Constant opacity}
From Fig.~\ref{fig:qrmsol} it is easy to discern what disc parameters will be able to prevent low-mass planets from migrating to the 
star. To test that hypothesis we ran numerical simulations of single planet formation with pebble accretion in both static and dynamic 
discs. We use the disc model from \citet{ida16}, which is based on the works of \citet{GL07} and \citet{Oka11}, for simplicity, but 
the analysis below can also be applied to more complex disc models e.g. \citet{KL10}. The disc is assumed to be in a steady state 
and the temperature and surface density are power laws of the distance. The best fit for the temperature profile for a solar-type star 
is given by \citep{GL07,ida16}

\begin{eqnarray}
 T_{\rm vis} &\simeq& 200 \alpha_3^{-1/5}\dot{M}_{*8}^{2/5}
\Bigl(\frac{r}{1\,{\rm AU}}\Bigr)^{-9/10}\; {\rm K}, \nonumber \\
T_{\rm irr} &\simeq& 150\Bigl(\frac{r}{1\,{\rm AU}}\Bigr)^{-3/7}\; {\rm K},
\label{eq:T_visirr}
\end{eqnarray}
where $\dot{M}_{*8} = \dot{M}_*/10^{-8}\,M_\odot\,{\rm yr}^{-1}$ and $\alpha_3 = \alpha/10^{-3}$. The inner region of the disc is 
viscously heated while the outer region is heated by the stellar flux. { For equation~(\ref{eq:T_visirr}) we assumed that 
the disc is vertically optically thin, but radially optically thick. In the case that the disc is so depleted that it is also 
optically thin in the radial direction, the temperature profile becomes $T \simeq 280 (r/1\,{\rm AU})^{-1/2}{\,\rm K}$ 
\citep{Hayashi81}. The radially optically thin condition occurs only when $\dot{M}_* \le 10^{-10}$~$M_\odot$~yr$^{-1}$. However, 
the condition for the disc to be radially optically thin only occurs at very low stellar accretion rate and accordingly a very low 
disc gas surface density. We do not consider this here because photoevaporation would destroy the disc before such low accretion 
rates can be attained \citep{A14}.}\\

From the above temperature relation the reduced scale height becomes
\begin{eqnarray}
h_{\rm vis} &\simeq& 0.027 \alpha_3^{-1/10}
\dot{M}_{*8}^{1/5}\Bigl(\frac{r}{1\,{\rm AU}}\Bigr)^{1/20},\nonumber \\
h_{\rm irr} &\simeq&0.024 \Bigl(\frac{r}{1\,{\rm AU}}\Bigr)^{2/7}.
\label{eq:h_visirr}
\end{eqnarray}
The surface density then follows from the steady state accretion. In the viscous region $d\ln \Sigma/d \ln r = -3/5$ and in the 
irradiative region $d\ln \Sigma/d \ln r = -15/14$. The boundary between the viscous and irradiation regimes given by $T_{\rm vis} = 
T_{\rm irr}$, which occurs at 
 \begin{equation}
r_{\rm vi} \simeq 1.8\alpha_3^{-14/33} \dot{M}_{*8}^{28/33}\, {\rm AU}.
\label{eq:r_vis_irr}
\end{equation}
{ The snow or ice line, which is the distance at which water ice condensates, is given by \citep{ida16}
\begin{equation}
r_{\rm snow} \sim \max(1.2 \alpha_3^{-2/9}\dot{M}_{*8}^{4/9}, 0.75)\quad {\rm AU}.
\end{equation}
} We employ this disc model and numerically simulate the evolution of a planet that forms through pebble accretion and migrates 
through the disc. We computed the evolutionary tracks according to \citet{Bitsch15}, which is slightly different from the method of 
\citet{ida16}. A summary of this method is given in the Appendix. { We have modified the code to employ the disc model 
discussed above; the pebble accretion model stays the same.} We first assume a constant accretion rate of $\dot{M}_{*8} = 
1$, corresponding to a disc age of 1~Myr, and set the inner edge of the disc at 0.1~AU { because it is unclear how reliable the 
employed disc model is closer to the Sun where MRI effects play an increasingly important role. For simplicity} we also set $\nu=\chi$ 
and $\gamma_{\rm eff}=7/5$ \citep{P11}. { Since we are interested in super-Earth formation and keeping the calculations feasible,} 
we place planetary seeds from 0.1~AU to 5~AU in succession for a range of initial values of $\alpha$ to determine the disc conditions 
that are required to prevent the planets from migrating to the star. { Placing planetary seeds at greater orbital distances would 
allow these seeds to grow to larger planetary cores because of the increased scale height of the disc. These planetary cores would 
then 
efficiently accrete gaseous envelopes and form gas giants \citep{I00}, hence our decision to place planetary seeds in the inner 
disc only.}\\

\begin{figure}[ht]
\resizebox{\hsize}{!}{\includegraphics[]{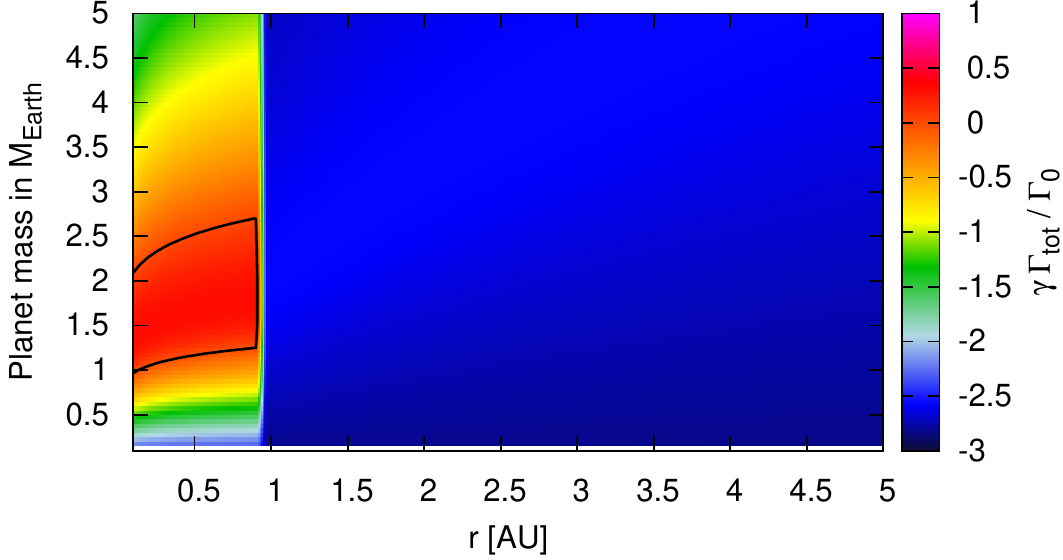}}
\caption{Torque map of the static disc with $\alpha=0.005$ and $q=9/10$ in the inner, viscously-heated part of the disc. Planetary 
masses inside the red region encompassed by the black line migrate outwards, indicating that planets can easily be trapped directly at 
the zero-torque locations (the black lines themselves). This indicates why planets only in a certain region of Fig.~\ref{fig:alpha90} 
can survive from inward migration. These are mainly the planets that are quite small (up to 2.5~M$_\oplus$) and in discs with higher 
viscosity.}
\label{fig:torque90}
\end{figure}

\begin{figure}[ht]
\resizebox{\hsize}{!}{\includegraphics[]{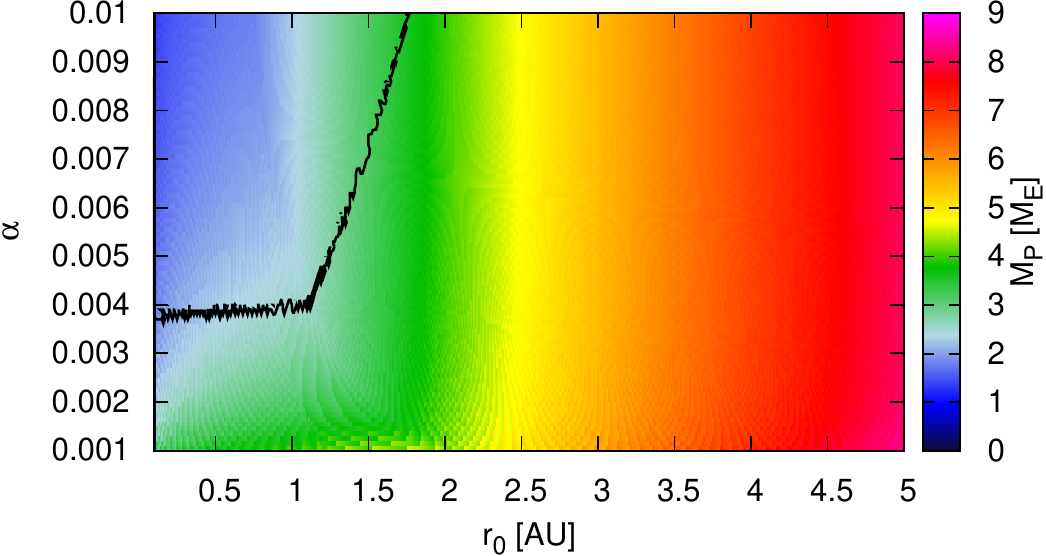}}
\caption{Final masses of the planets as a function of their initial position, $r_0$, and of the $\alpha$ viscosity 
parameter in a static disc with $q=9/10$ in the inner, viscous-heated part. Planets with a given initial $\alpha$ and $r_0$ that are 
above and to the left of the black line (top left part of the plot) end up on orbits farther than 0.1~AU, which marks the inner edge 
of the disc. All other planets are lost due to inward migration because they are too massive to be contained in the region of outward 
migration.}
\label{fig:alpha90}
\end{figure}

\begin{figure}[ht]
\resizebox{\hsize}{!}{\includegraphics[]{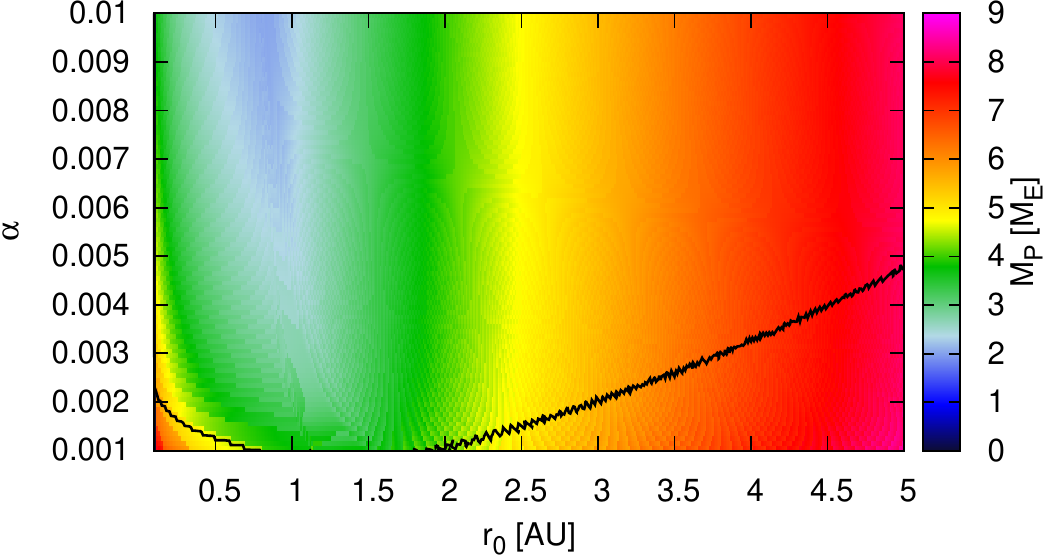}}
\caption{Identical to Fig.~\ref{fig:alpha90} but now the temperature gradient is 6/5 rather than 9/10 in the inner, viscous-heated 
part. Note that the threshold value of $\alpha$ is factor of a few lower than in the previous case, and that the volume of phase space 
where planets are trapped is also larger.}
\label{fig:alpha120}
\end{figure}

In Fig.~\ref{fig:torque90} we plot the normalised torque as a function of distance to the Sun and planetary mass for our chosen, 
static disc model. The red region encompassed by the black line is where outward migration occurs, though the maximum mass for 
outward migation is always below 3~M$_\oplus$. { The region of outward migration is smaller than that found in \citet{B15} because 
of the different disc model that we have chosen. In \citet{B15} the transition between the viscous and irradiative regimes of the disc 
is smooth but with a complicated structure, while in our simplified model the change is abrupt: the temperature and surface gradients 
alter suddenly at $r_{\rm vi}$, which is generally closer than 2~AU to the star.}\\

When we perform single planet pebble accretion simulations with migration for a range of initial distances to the Sun and viscosity, 
we end up with the result displayed in Fig.~\ref{fig:alpha90}. The black line indicates the critical value of $\alpha$ above which the 
planets end up on orbits farther than 0.1~AU from the Sun; all other planets are lost to the star. It is clear that for a temperature 
slope of $9/10$ a critical $\alpha \sim 4 \times 10^{-3}$ is necessary to prevent the planets from migrating to the star.\\

Increasing the temperature gradient will decrease the critical value of $\alpha$ that is required to trap planets far from the star, 
while it also enhances the overall phase space where this trapping becomes possible. { This is shown in Fig.~\ref{fig:alpha120}, 
which is identical to Fig.~\ref{fig:alpha90} but now the temperature gradient is increased to $q=6/5$ in the viscous-heated part. As 
mentioned earlier, the typical range of temperature gradient in the viscous region is $0.85\lesssim q \lesssim 1.2$, so 
Figs.~\ref{fig:alpha90} and \ref{fig:alpha120} represent approximate end member states. In Fig.~\ref{fig:alpha120} we assumed that the 
whole viscous region of the disc has this steep slope, which is not observed in the simulations of \citet{B15}. Thus the outcome 
should be treated as an extreme case. With the steeper temperature gradient the area of phase space in Fig.~\ref{fig:alpha120}, where 
planet trapping now becomes possible, has increased substantially. This enlarged region is caused by the elevated strength of the 
corotation torque with temperature gradient, as shown in Fig.~\ref{fig:pqtorque} and discussed in Section~1. In addition, a change in 
the temperature gradient also results in an opposite variation in the surface density gradient, which becomes shallower as the 
temperature gradient grows steeper. The shallower surface density gradient also supports outward migration.} \\

Thus, it seems that it is certainly possible to prevent low-mass planets from reaching the star under the right conditions, provided 
that a threshold value of $\alpha$ is surpassed; this critical value is a strong function of the temperature gradient, however.

\subsection{Constant opacity with time evolution}
Circumstellar discs evolve with time, { with the surface density and temperature readily decreasing with time while $\alpha$ and 
$q$ are often assumed to remain constant}. The cases displayed above are for a disc with a fixed value of $\dot{M}_*$ while in reality 
this quantity decreases with time according to \citep{H98}

\begin{equation}
\log \dot{M}_{*8} = -1.4\log\Bigl(\frac{t}{{\rm Myr}}+0.1\Bigr).
\label{eq:mdot}
\end{equation}
{ The extra factor of 0.1 was added by \citet{B15} to avoid the singularity when $t=0$. The relation in equation~(\ref{eq:mdot}) 
is based on observations and it is not very well constrained, but it is widely used by the community. From 
equation~(\ref{eq:mdot}) and a fixed value of $\alpha$ and a temperature relation one can compute the surface density through the 
usage of equation~(\ref{eq:dotmstar}).}\\

Very young discs have an accretion rate upwards of $10^{-7} M_\odot$~yr$^{-1}$, which results in a very hot inner disc with a large 
aspect ratio. Any planets that form early in this disc quickly reach a large isolation mass and are (eventually) trapped in the region 
of outward migration. However, as the disc evolves with time, the region of outward migration shrinks, the planets are released from 
it and they fall into the central star \citep{Bitsch15}. In this study we only focus on cases where the disc age is 1~Myr or older; 
the maximum disc age is 3~Myr.\\

\begin{figure}[ht]
\resizebox{\hsize}{!}{\includegraphics[]{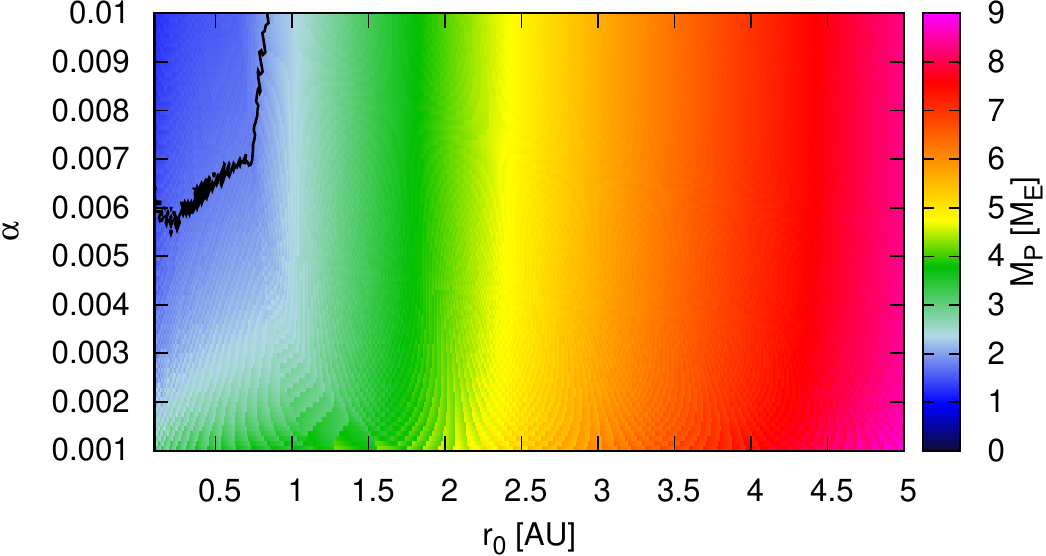}}
\resizebox{\hsize}{!}{\includegraphics[]{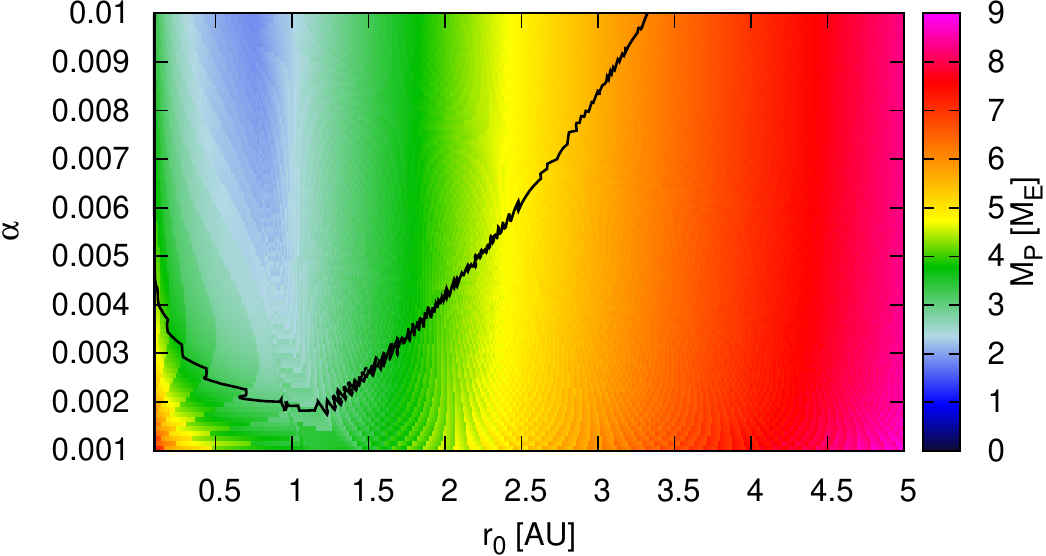}}
\caption{Final masses of the planets' position as a function of their initial position $r_0$ and of the $\alpha$ parameter, but now 
the disc evolves with time. The initial disc age is 1~Myr. In the inner parts of the disc the temperature gradient is $9/10$ (top) and 
$6/5$ (bottom). Planets with a given $\alpha$ and $r_0$ that are above the black line (top left part of the plot) end up on orbits 
larger than 0.1~AU. All other planets are lost to the star.}
\label{fig:alpha090120Myr1}
\end{figure}

\begin{figure}[ht]
\resizebox{\hsize}{!}{\includegraphics[]{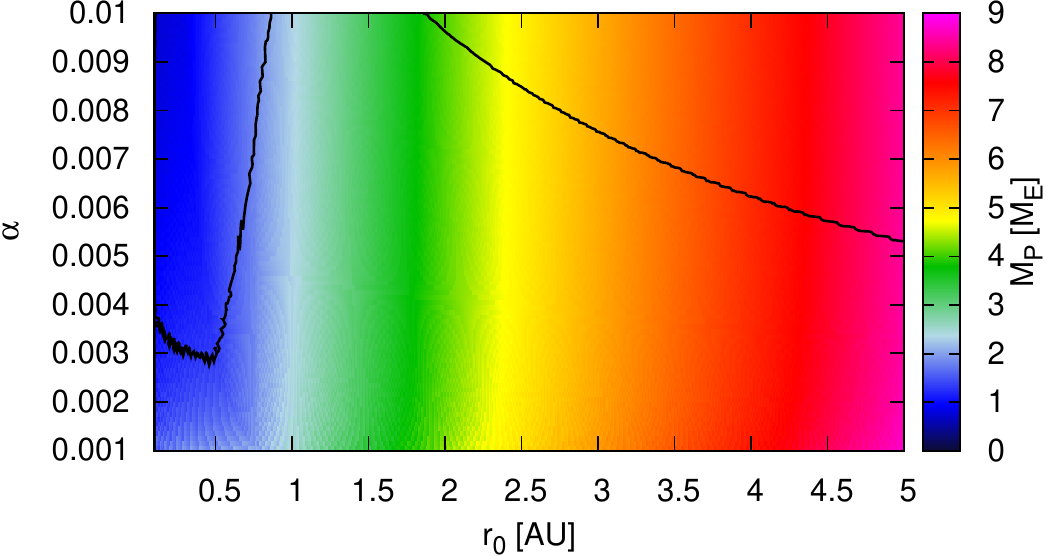}}
\resizebox{\hsize}{!}{\includegraphics[]{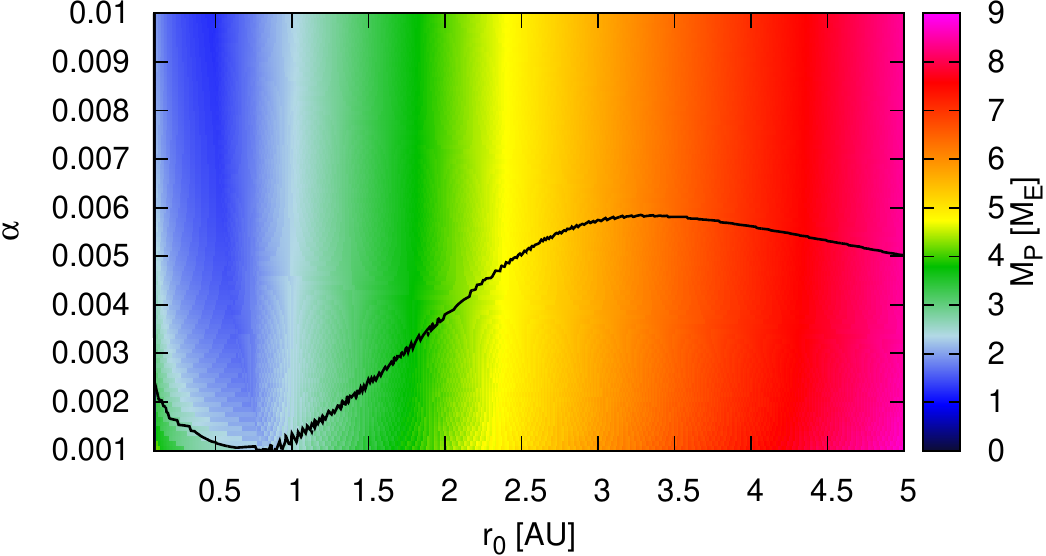}}
\caption{Same as Fig.~\ref{fig:alpha090120Myr1} but now the initial disc age is 2~Myr.}
\label{fig:alpha090120Myr2}
\end{figure}

Figure~\ref{fig:alpha090120Myr1} shows the region where planets may be trapped farther than 0.1~AU as a function of $\alpha$ and 
initial formation location for an evolving disc. In the top panel the temperature gradient in the inner disc is $9/10$ 
while it is $6/5$ in the bottom panel. It is clear that the disc evolution increases the critical value of $\alpha$ and also shrinks 
the total area of the plot where the planets may be trapped.\\

Increasing the initial disc lifetime to 2~Myr yields similar results, which are shown in Fig.~\ref{fig:alpha090120Myr2}. The final 
masses of trapped planets are generally lower than for the younger disc because the aspect ratio is now diminished (the disc is 
cooler). However, the planets can be trapped more easily in the region of outward migration as the region itself barely evolves beyond 
this point as the disc cools. The region where planets survive in the disc is therefore larger than it is for the younger (and warmer) 
disc. Additionally, the region in $r_0$-$\alpha$ space where planets survive extends all the way to the outer disc. These planets 
actually have not had enough time to migrate all the way to the central star as the disc dissipates at 3~Myr. This can best be seen in 
the top panel of Fig.~\ref{fig:alpha090120Myr2} where the two regions of surviving planets appear separated, while in the bottom 
panel, where the temperature gradient is steeper, the variation in the critical value of $\alpha$ as a function of $r_0$ is less 
pronounced.

\subsection{Variable opacity}
Thus far we have avoided discussing the influence of the disc's opacity because it adds another dimension to the problem, increasing 
its complexity. { Unfortunately there are various popular opacity laws in existence { \citep{Bell94,Bell97,Sem03}}, so model choice 
will affect the outcome. Here we shall use the opacity profile of \citet{Bell94} because this was used by \citet{B15} and it makes for 
easier comparison with the full disc model discussed in the next subsection.}\\

{ The temperature structure of the protoplanetary disc in the viscously dominated part is determined by the balance between heating 
and cooling, where the cooling function is inversely proportional to the opacity of the disc. A steep gradient in the opacity function 
will thus result in a steep gradient in the temperature profile. However, the cooling also depends on the density of the 
protoplanetary disc, where a higher density blocks outgoing radiation and decreases the cooling rate. The heating, on the other hand, 
also depends on the density and the viscosity. The only way to safely determine the temperature structure of protoplanetary discs are 
thus 2D simulations, including heating and cooling \citep{B15}. However, these simulations are beyond the scope of this work because 
we want to disentangle the interplay between the different factors responsible for growth and migration of super-Earth planets. For 
this purpose we rely on simple power-law discs, which gives us an easier understanding of the interactions of the parameters, even 
if the temperature and opacity gradients may not everywhere be fully consistent.\\}

Put more simply, the role of the opacity is to change the value of $\chi$ and thereby the cooling rate of the disc, effective ratio of 
the specific heats of the disc, $\gamma_{\rm eff}$, and the value of $p_{\rm \chi}$ versus $p_{\nu}$ \citep{P11}. { Here we want to 
show one case of how the opacity likely affects the outcome, keeping in mind there is a larger variety of outcomes depending on the 
opacity relation that is employed and assumptions about how the opacity scales with temperature and density \citep{GL07}.} \\

\begin{figure}[ht]
\resizebox{\hsize}{!}{\includegraphics[]{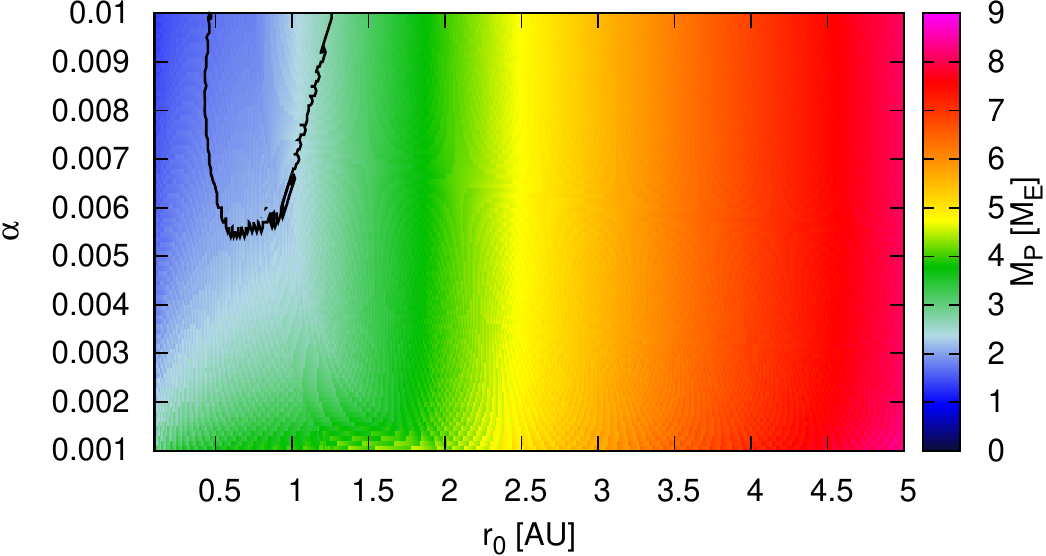}}
\resizebox{\hsize}{!}{\includegraphics[]{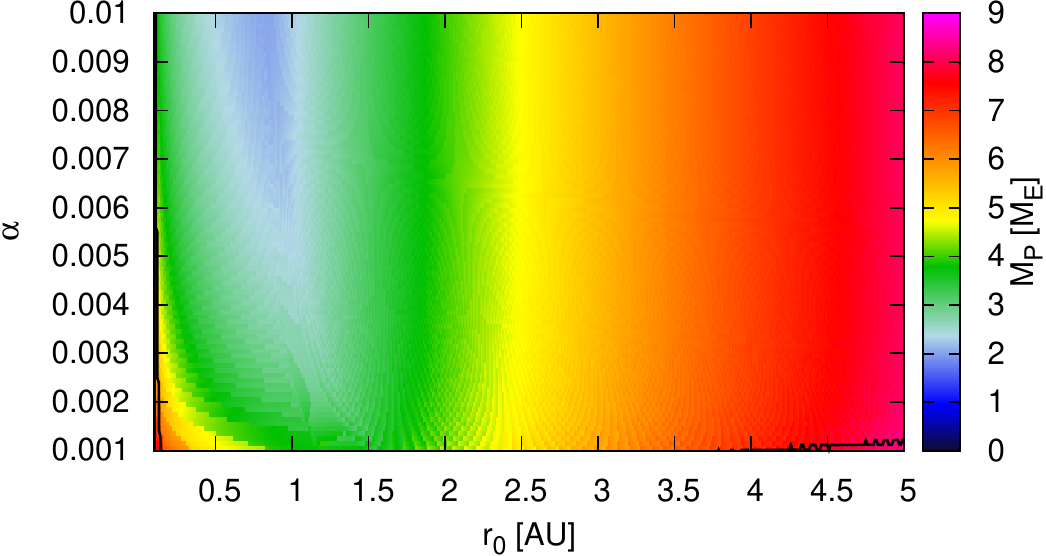}}
\caption{Final masses of the planets' position as a function of their initial position $r_0$ and of the $\alpha$ parameter for a 
static disc with an age of 1~Myr. The opacity is allowed to vary with temperature and density of the disc according to the 
prescription of \citet{Bell94}. In the inner parts of the disc the temperature gradient is 9/10 (top) and 6/5 (bottom). { In the 
bottom panel all the planets are saved.}}
\label{fig:alpha090120kappa}
\end{figure}

In Fig.~\ref{fig:alpha090120kappa} we employ a static disc with $\dot{M}_{*8}=1$, corresponding to an age of 1~Myr, but the opacity is 
allowed to vary. Specifically we use the opacity relations of \citet{Bell94}. It is clear that allowing the opacity to vary is not of 
much help in the top panel, for which the temperature gradient in the inner disc is 9/10. The critical $\alpha$ is higher than in the 
static disc with constant opacity, and the region encompassed by the black line is smaller. However, when the temperature gradient is 
6/5, as in the bottom panel, the critical $\alpha$ is reduced below $10^{-3}$ and the region encompassed by the black line is much 
greater than in the static disc of Fig.~\ref{fig:alpha120}.

\subsection{Variable opacity with time evolution}
\begin{figure}[ht]
\resizebox{\hsize}{!}{\includegraphics[]{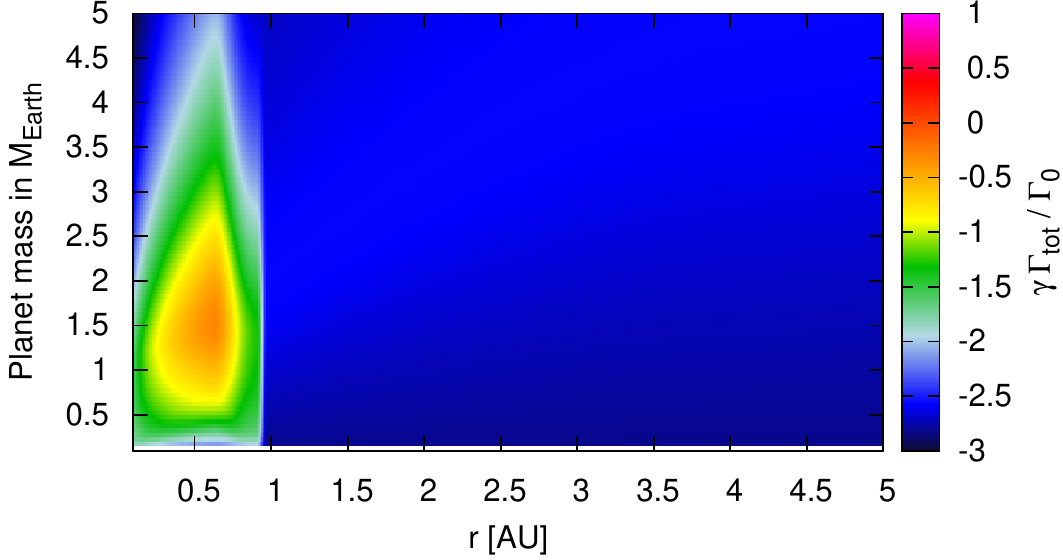}}
\resizebox{\hsize}{!}{\includegraphics[]{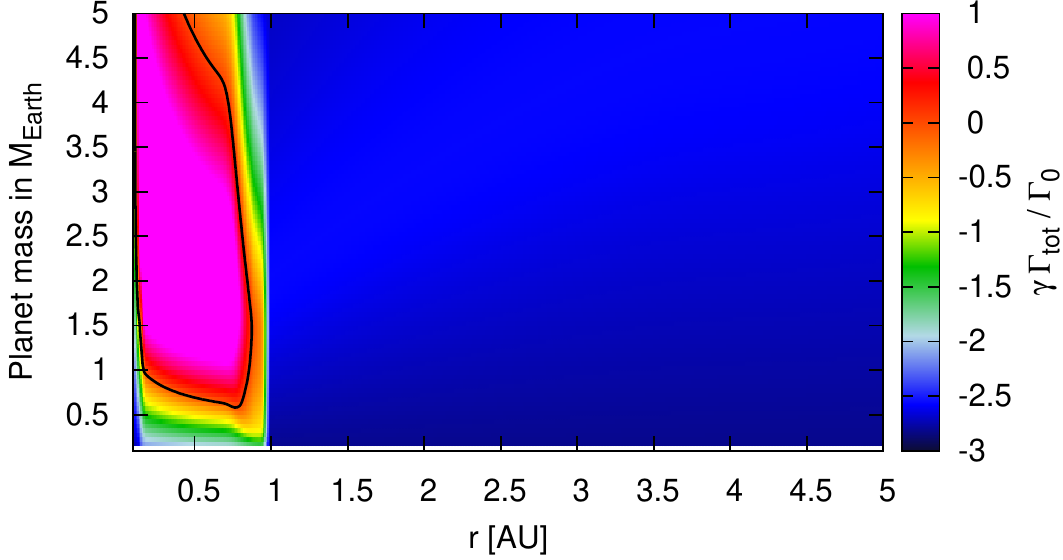}}
\caption{Same as Fig.~\ref{fig:torque90} but now the effect of the opacity is included. This snapshot is at 1~Myr and $\alpha=0.005$. 
In the top panel the temperature gradient in the viscous region is 9/10 and 6/5 for the bottom panel. There is no region of outward 
migration in the top panel, which is different from that of Fig.~\ref{fig:torque90}. This is caused by the opacity varying the ratio 
$\chi/\nu$, and thus $p_\chi/p_\nu$, which in turn affects the strength of the corotation torque.}
\label{fig:torque090120kappaevolve}
\end{figure}

\begin{figure}[ht]
\resizebox{\hsize}{!}{\includegraphics[]{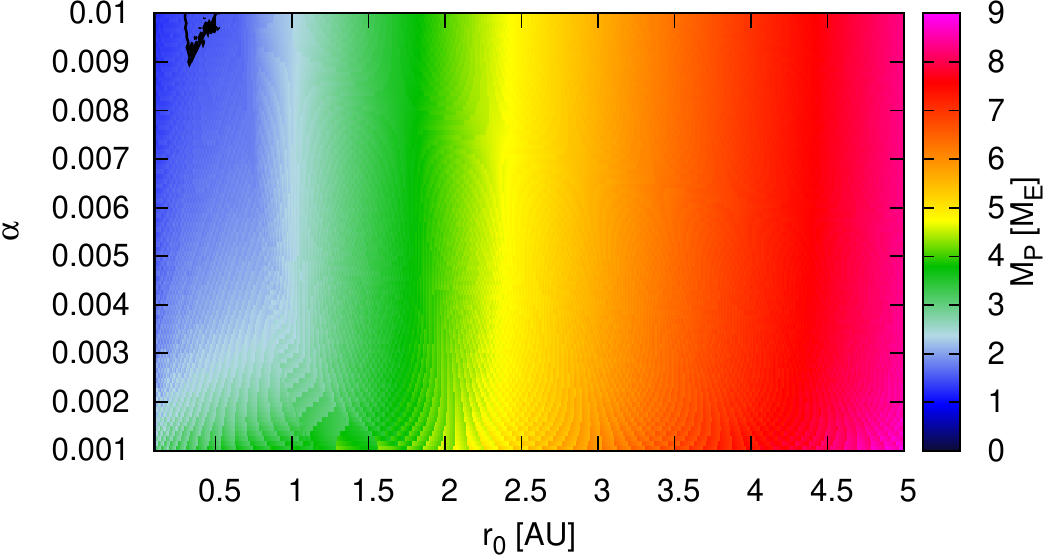}}
\resizebox{\hsize}{!}{\includegraphics[]{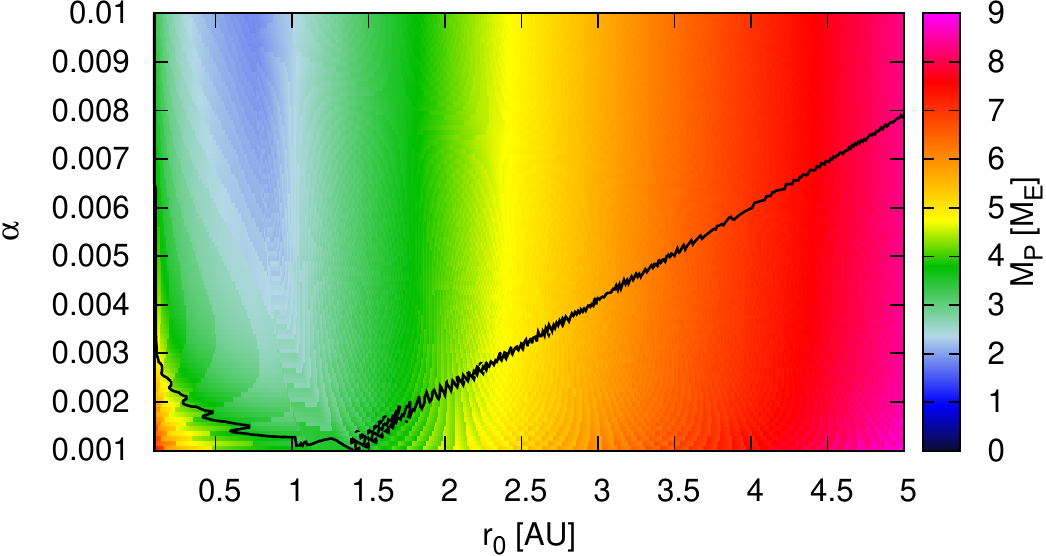}}
\caption{Same as Fig.~\ref{fig:alpha090120kappa} but now the disc is evolving with time. The initial disc age is 1~Myr. Only a steep 
temperature gradient allows for outward migration.}
\label{fig:alpha090120kappaevolve}
\end{figure}
{ When we add time evolution to the simplified disc with a variable opacity but a static temperature gradient, the migration map is 
depicted in Fig.~\ref{fig:torque090120kappaevolve}. It is clear that this is substantially different than what was displayed in 
Fig.~\ref{fig:torque90}. Indeed, when $q=9/10$ (top panel), there is no region of outward migration, precisely because the opacity 
changes $\chi/\nu$ which potentially lowers the effect of the corotation torque. In the bottom panel, where $q=6/5$, the region of 
outward migration is substantial, as is expected.} The final masses and area where planets may be trapped in the outward migration 
region is displayed in Fig.~\ref{fig:alpha090120kappaevolve}. { Note that, in the top panel, almost no planets are saved from 
falling into the star, while in the bottom panel most planets are fine.}\\

In summary, the inclusion of a varying opacity only appears to be beneficial for steep temperature profiles, but is a hindrance for 
shallow temperature profiles because outward migration seems to be more sensitive to the value of $\chi/\nu$.

\subsection{Full model}
Here we briefly touch upon using a full disc model such as the one in \citet{B14}.\\

\begin{figure}[ht]
\resizebox{\hsize}{!}{\includegraphics[]{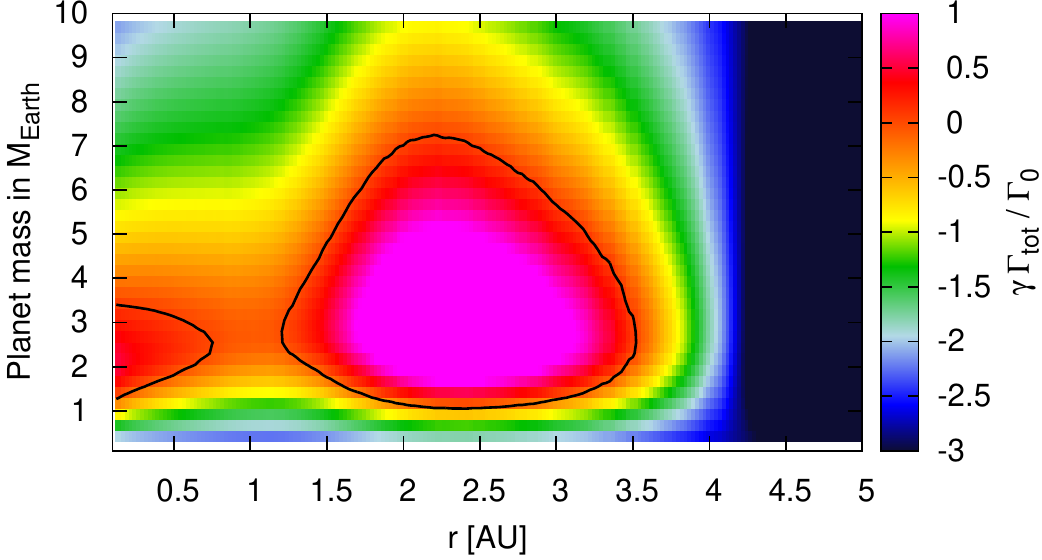}}
\caption{Same as Fig.~\ref{fig:torque90} but now employing the full disc model of \citep{B15}. The disc age is 1~Myr, 
$\alpha=0.0054$ and the opacity relation of \citet{Bell94} was used. Note the two regions of outward migration.}
\label{fig:torquefull}
\end{figure}

\begin{figure}[ht]
\resizebox{\hsize}{!}{\includegraphics[]{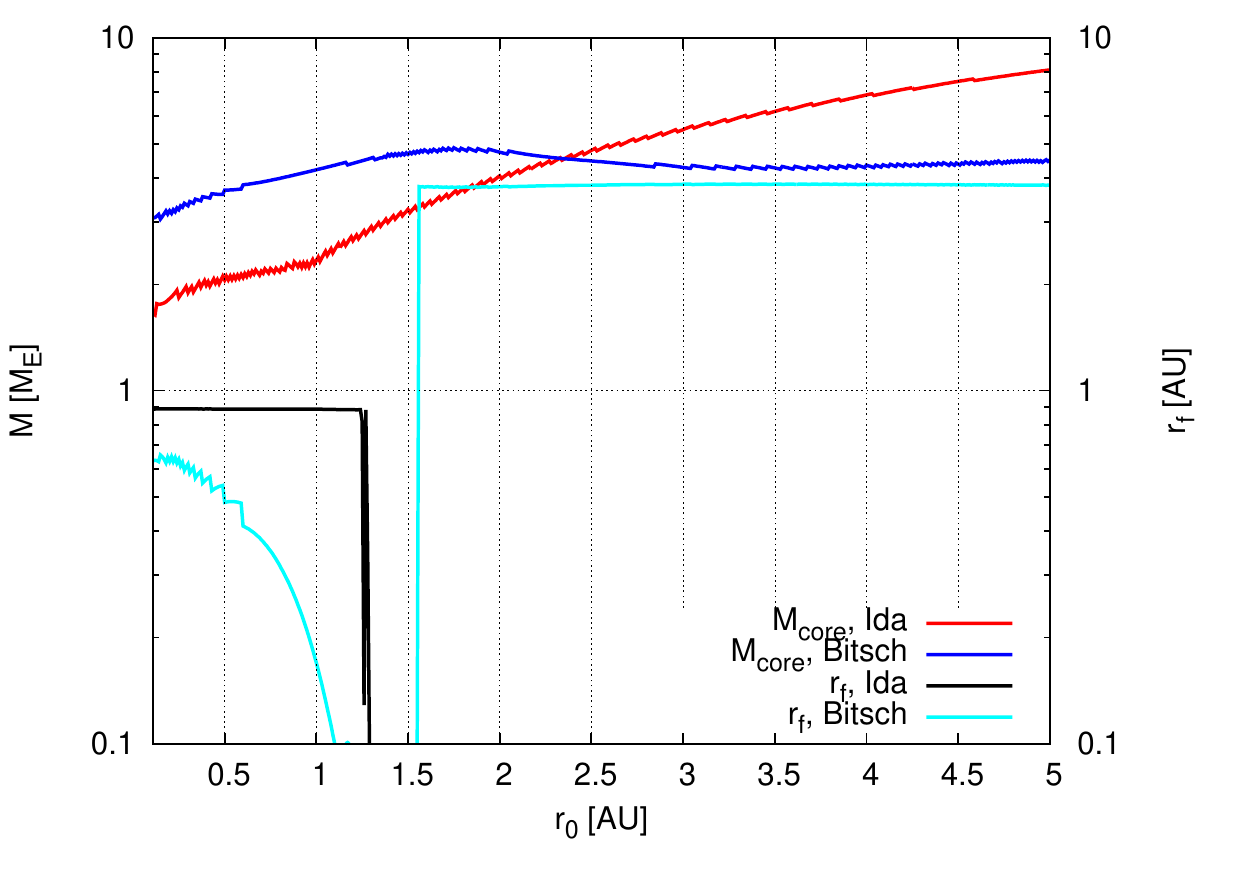}}
\caption{Comparison of the final positions and masses of planets as a function of their initial birth distance for this single full 
disc model and the simplified power law model. For the simplified disc model $q=9/10$.}
\label{fig:slopecompare}
\end{figure}

{ In Fig.~\ref{fig:torquefull} the migration map for the full disc model of \citet{B15} is displayed a the point when the disc age 
is 1~Myr. Note that there are two regions of outward migration: one close to the star and one at moderate distance. These are caused 
by transitions in the opacity. The outer region was not present in the simplified model which assumes the disc is fully irradiative 
beyond $r_{\rm vi}$, in contrast to the disc model of \cite{B15}, where a shadowed region exists for $r>r_{\rm ice}$ allowing outward 
migration. The inner migration region is caused by the silicate line and the outer one is cause by the water ice line. \\

Unfortunately we cannot display a figure similar to Fig.~\ref{fig:alpha90} because the 2D disc model simulations performed so 
far used $\alpha=0.054$ { to model the viscous heating}. Computing the evolution of discs with varying $\alpha$ is computationally 
too expensive. Thus the migration map should be used as an indication of the outcome. Instead, in Fig.~\ref{fig:slopecompare} we 
present the final positions and masses of planets as a function of their initial birth distance for this single full disc model and 
the simplified power law model we used earlier for the same $\alpha$ and with the viscous temperature gradient set to 9/10.\\

What is clear is that planets in the power law disc only survive in the inner part, where the temperature slope is steep enough to 
trigger outward migration. In between the two regions of outward migration in the disc model of \cite{B15}, the planets grow too 
massive (about 3.5 $M_\oplus$) to be contained in the inner region of outward migration and thus move all the way to the inner disc 
edge. Note that the final migration map at $t=3$~Myr (not shown) allows only even smaller planets to stay in the region of outward 
migration compared to Fig.~\ref{fig:torquefull}.\\

Thus the full disc model certainly helps in stalling planets farther from the star. The main reason is that there are two regions of 
outward migration (see Fig.~\ref{fig:torquefull}). The first is inside 0.5~AU and is the result of viscous heating of the disc. The 
second region between 2~AU and 4~AU is the result of shadowing of the disc due to a drop in the opacity profile. In the shadowed 
region $h$ shrinks with distance and thus the temperature gradient is sufficiently steep for the corotation torque to dominate the 
migration.}

\section{Discussion}
There are several topics that require further discussion but which go beyond the scope of the paper and could be the subject of 
future studies.\\

{ In our models the temperature profile does not evolve in the outer parts, beyond $r_{\rm vi}$, and thus the aspect ratio there is 
constant in time. Only the surface density decreases because $\dot{M}_{*8}$ decreases. However, this is unlikely to affect the overall 
result because in the outer part of the disc type I migration is almost always inward (see Section 2.2). \\

The evolution of the inner part of the disc is more complicated than we assumed here. Even though the temperature and surface density 
gradients remain the same, and the steep temperature gradient allows for outward migration, the temperature itself decreases, and thus 
the disc's aspect ratio also shrinks. Since the temperature in the viscous region declines, eventually stellar heating will take 
over and the irradiative regime moves inwards. This implies that the zero-torque region must also shift sunwards because the amount 
of viscous heating subsides. Thus all planets then follow the zero-torque line and only the most massive ones would pop out of 
the outward migration region and subsequently fall into the star.\\

Is there a way to prevent this outcome? One way could be to suppose that the disc dissipates very rapidly once the planets have 
halted in the outward migration region. In order for this to work, the disc's dissipation time scale would have to be shorter than 
the migration time scale. The migration time is normally a few hundred thousand years for an Earth-mass planet, which is much shorter 
than the dissipation time of the disc, especially when the planet is at a few AU. { Only when the stellar accretion rate drops 
$10^{-10}$~$M_\odot$~yr$^{-1}$ is the clearing of the disc by photo evaporation much faster \citep{B15}, and comparable to the 
migration time scale. In summary, we think that this scenario is implausible most of the time.\\}

{ In this work, we focused on planets which grow to their pebble isolation masses. However, it is also feasible to imagine that 
the final masses of planets are $m_p < m_{\rm iso}$. Such planets could form late. Alternatively, they could appear after the pebble 
flux dries up because planets farther out have reached their isolation mass before the inner ones have, and thereby prevent them from 
growing any further. Such planets could also be grown by planetesimal accretion, whose `planetsimal' isolation mass \citep{JJL87} is 
generally lower than the pebble isolation mass. In this case the same issues apply as for a planet that has reached the isolation 
mass. Whether or not such a planet remains trapped in the outward migration region depends on the precise evolution and properties of 
the disc. It is not possible to determine this {\it a priori}.\\}

The exact formation pathway of super-Earths is still a mystery. Do they form in-situ or via migration? In one of the earliest 
studies of its kind, \citet{OI9} simulated rocky planet formation around an M-dwarf star starting with a disc of planetesimals and 
evolving it all the way past a possibly giant impact phase. They studied two end cases: one with the full strength of type I migration 
included and one where the migration was artificially slowed down. They discovered that, in the slow migration case, almost all 
planets ended up in a multiple resonant chain. With full migration they only formed a few planets close to the star, generally fewer 
than with the slower migration.\\

Later work by \citet{HM13} combined N-body and Monte Carlo simulations. They suggest that in-situ formation of super-Earth systems is 
possible and provides a reasonably good match to existing {\it Kepler} data. On the other hand, \citet{RC14} argue against in-situ 
formation and suggest that all close-in super-Earth planets must have reached their current orbits as a result of migration. 
Recently \citet{O15} studied the formation pathway of super-Earths in discs that do not allow outward migration. In their study, they 
let planetary embryos migrate towards the inner edge of the disc where they collided and formed bigger objects. The resulting 
planetary systems were basically all tightly packed and with a hierarchical mass order (the most massive planet was closest to the 
central star) in contradiction to observations. This suggests that planet migration in a disc with no zones of outward migration may 
not be able to explain the distribution of observed super-Earth systems. Thus a change in the migration mechanism may be necessary to 
explain the observed distribution of super-Earths. The results of \citet{C14} and \citet{CN14} are somewhat more promising, but all of 
these studies rely on { planetesimal accretion and on} migration to pile up the planets close to the central star. { The 
formation of multiple planets undergoing pebble accretion is expected to have a great variety of outcomes, with the mass gradient of 
the planets sensitively depending on the temperature gradient of the disc \citep{ida16}. A full N-body study is beyond the scope of 
this paper.} \\

Zones of outward migration in the protoplanetary disc can be caused by strong enough radial gradients in temperature, as we studied 
here, or by an inversion of the radial surface density profile, where the surface density increases with orbital distance 
\citep{Masset06}. Those changes in the disc profile can be achieved due to photoevaporation, which carves a hole in the inner disc, 
stopping planet migration, even for gas giants \citep{AP12}. However, these changes of the surface density gradients only appear in 
the very late stages of the disc evolution and are typically farther than 1~AU, which severely limits the ability of this mechanism to 
halt planet migration and prevent super-Earth planets from reaching the central star.\\

{ Another topic that requires discussion is the effect of whether or not the pebble isolation mass depends on $\alpha$; this was 
alluded to in Section~2.3. Suppose that $m_{\rm iso} \propto \alpha_3^\beta$ { and assume this relation is valid for $\alpha_3 
\gtrsim 0.01$. We repeat that $\alpha_3=\alpha/10^{-3}$ and we explore the scaling law around this value.} Then we have 

\begin{equation}
h_{\rm crit-\alpha} = \frac{\pi\sqrt{3}}{128}p_\nu^2\alpha_3^{1-3/2\beta}.
\label{eq:hcrit2}
\end{equation}
We assume $\beta \ll 1$. { This choice is justified because \citet{Lam14} observed no appreciable difference in the isolation mass 
when they varied $\alpha$ by a factor of five. In an independent study \citet{Zhu14} conclude that in inviscid discs the estimate of 
the isolation mass is no more than a factor of two lower than the nominal value; a value $\beta \gg 1$ appears inconsistent with these 
previous works.}\\

As an example we consider $\beta=1/5$ then $h_{\rm crit} \propto \alpha_3^{7/10}$. As $\alpha_3$ is varied from 0.1 to 10 the value of 
$h_{\rm crit}$ varies by as much as a factor five from the case { where the isolation mass has} no $\alpha$-dependence. 
{ Dividing the modified value of $h_{\rm crit-\alpha}$ in equation~(\ref{eq:hcrit2}) by the original value $h_{\rm crit}$ in 
equation~(\ref{eq:hcrit}) yields $h_{\rm crit-\alpha}/h_{\rm crit}=\alpha_3^{-3/2\beta}$. Thus the modified value $h_{\rm 
crit-\alpha}$ is higher than the original value in equation~(\ref{eq:hcrit})} for $\alpha_3<1$ and lower when $\alpha_3>1$. { This 
$\alpha$ dependence hinders trapping planets in the outward migration region: when $\alpha_3>1$ the pebble isolation mass becomes 
larger than without an $\alpha$ dependence, and the trapping needs to occur at lower $h$ (see Fig.~\ref{fig:qrmsol})}. When 
$\alpha_3<1$ { in theory} the trapping should be more efficient because the zero-torque line is crossed at higher $h$. { In 
Fig.~\ref{fig:alpha090120kappaevolvealpha} we show the outcome of simple pebble accretion simulations in an evolving disc with varying 
opacity but now the pebble isolation mass scales as $m_{\rm iso} \propto \alpha_3^{1/5}$. In the top panel no planets survive, which 
is the same outcome as in Fig.~\ref{fig:alpha090120kappaevolve}. In the bottom panel there is a clear dependence on $\alpha$ and 
actually the larger pebble isolation mass for larger $\alpha$ makes trapping more difficult in the evolving disc compared to the case 
with no alpha dependence (Fig.\ref{fig:alpha090120kappaevolve}). Note also the slightly different colour scale and maximum isolation 
mass, confirming our expectations.}\\

If there is no $\alpha$ dependence on $m_{\rm iso}$ then it becomes nearly impossible to trap low-mass planets in the simplified model 
that we have considered if the temperature gradient is shallow. A viscous region with a steeper temperature gradient works much 
better, as does the full model because of the large region of outward migration near 3~AU. On the other hand, if the weak $\alpha$ 
dependence is confirmed in future studies, then discs having $\alpha<10^{-3}$ { are still not expected to be able to trap planets 
far from the star}.\\

\begin{figure}[ht]
\resizebox{\hsize}{!}{\includegraphics[]{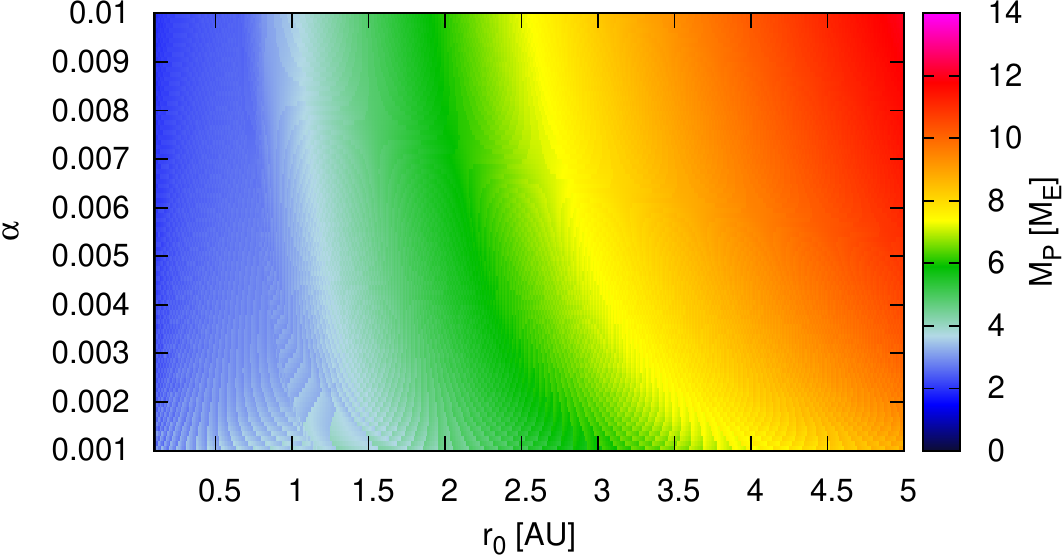}}
\resizebox{\hsize}{!}{\includegraphics[]{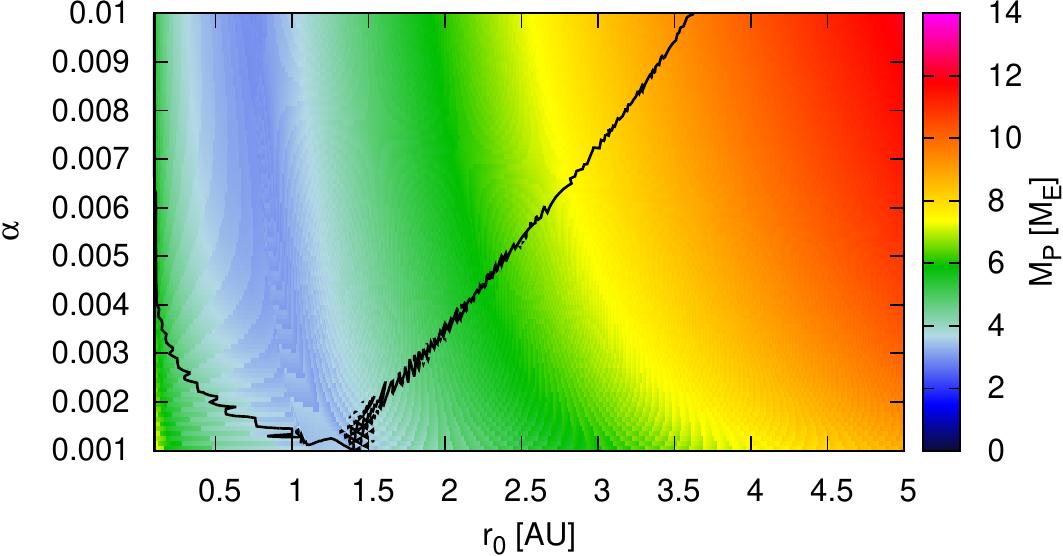}}
\caption{Same as Fig.~\ref{fig:alpha090120kappaevolve} but now the pebble isolation mass scales weakly with $\alpha$ as $m_{\rm iso} 
\propto \alpha_3^{1/5}$. The initial disc age is 1~Myr. Once again only a steep temperature gradient allows for outward migration and 
trapping planets far from the star.}
\label{fig:alpha090120kappaevolvealpha}
\end{figure}

A recent study of the global evolution of protoplanetary discs indicates that the midplane is laminar, with $\alpha$ likely 
to be lower than $10^{-3}$ \citep{Bai16}, which may not be fatal for planet migration if the pebble isolation mass had some $\alpha$ 
dependence. These newer models also do not appear to agree very well with traditional constant $\alpha$ disc models driven by 
isotropic MRI turbulence, however. That said, much information is still to be had about planet migration with the more traditional 
approach and models that we have adopted here. An alternative model, relying on disc winds \citep{Suz16} could be explored in the 
future. In addition, we are not yet aware of a full treatment of the torques on the planets from such a disc.\\}

{ Another possibility for trapping planets would be the inner edge of the dead zone, where changes in viscosity cause a 
positive gradient in the surface density, blocking the influx of dust and also the inward motion of planets \citep{F17,Masset06}. 
These changes in viscosity are caused by changes in the ionisation fraction of the disc, which is a function of the disc's density. 
However, as discs evolve, so will the inner edge of the dead zones, washing away the gradient of the surface density, releasing 
trapped planets \citep{B14}, which is why here we focus instead on trapping planets purely by the corotation torque instead.\\}

Unless there is an as of yet unknown and undiscovered disc mechanism that acts on migrating planets, the entropy driven corotation 
torque is the only way of stopping, and possibly reversing, the inward migration of super-Earth mass planets. For this to happen, the 
disc profile needs to have a radial temperature slope steeper than 0.87 and a viscosity larger than a few promille. Lower temperature 
slopes and viscosities in discs will only allow inward migration, leading to the { problematic} super-Earth formation scenario 
from \cite{O15}. It is not guaranteed that steeper temperature slopes and higher viscosities in discs will solve the formation mystery 
of super-Earths, but it should be investigated in future studies, especially those relying on pebble accretion.

\section{Summary and conclusions}
We investigated the conditions that are required for low-mass planets to execute outward migration in steady-state protoplanetary 
discs. This outward migration is accomplished by the horseshoe torque, which is a complex function of the mass, scale height, 
viscosity and thermal diffusivity of the disc. We restricted ourselves mostly to the disc model employed by \citet{ida16} because of 
its simplicity, but we allowed the temperature gradient and opacity to vary. We ultimately conclude the following:

\begin{itemize}
 \item the corotation torque is at a maximum when the thermal diffusion constant $\chi=\frac{9}{4}\nu$;
 \item the negative temperature gradient needs to exceed approximately 0.87 for outward 
migration to occur;
\item there exists a critical value of the disc scale height, $h_{\rm crit}$, below which planets at the pebble isolation mass 
are trapped in the outward migration region, preventing them from reaching the central star;
\item planets at the pebble isolation mass can be prevented from reaching the star as long as $\alpha$ exceeds a critical value that 
depends on the temperature profile and age of the disc. A typical rule of thumb is that $\alpha_{\rm crit} \gtrsim 4 \times 10^{-3}$;
\item opacity variations in the disc do not necessarily increase the region of phase space where planets can be trapped, but it may 
create additional regions where outward migration becomes possible.
\end{itemize}

{ We predict that systems with a high number of low-mass planets, such as TRAPPIST-1 \citep{Gil17}, whose planets may be in a 
multiple resonant chain, underwent slower migration than those with just a few planets \citep{OI9}. We attribute the variation in 
migration speed from one system to the next to a different temperature gradient and opacity in the viscous-heated part of 
their circumstellar discs.}\\

{\footnotesize 
The authors thank Michiel Lambrechts for a positive and constructive review that substantially improved this paper. RB is grateful for 
financial support from JSPS KAKENHI (16K17662).}

\section{Appendix A}
Here we briefly summarise our implentation of pebble accretion.\\

The time scale of growing planetary seeds from planetary embryos all the way to planetary cores can be greatly enhanced by considering 
the accretion of leftover pebbles in the disc \citep{Ormel10,LJ12}. In contrast to planetesimals, small pebbles feel gas drag and 
migrate through the disc. If these pebbles enter within the Hill sphere of the planet, they spiral towards the planet due to their loss 
of angular momentum caused by the friction with the gas and are subsequently accreted.\\

Very small bodies ($100-200$ km in diameter) accrete pebbles in the inefficient Bondi accretion branch \citet{LJ12}, where the 
accretion time scale is quite long (depending on the disc structure this can be of the range of 1~Myr) until the pebble transition 
mass is reached
\begin{equation}
\label{eq:Mtrans}
 M_{\rm t} = \frac{\sqrt{3}}{3} \frac{(\eta v_{\rm K})^3}{G \Omega_{\rm K}} \ ,
\end{equation}
where $G$ is the gravitational constant, $v_{\rm K}= \Omega_{\rm K} r$, and
\begin{equation}
\label{eq:eta}
 \eta = - \frac{1}{2} \frac{\partial \ln P}{\partial \ln r}h^2.
\end{equation}
Here, $\partial \ln P / \partial \ln r$ is the radial pressure gradient in the disc; generally $\partial \ln P / \partial \ln 
r=3-\frac{1}{2}q$. These masses are typically in the range of 0.01~$M_\oplus$ and it marks the mass at which planets can accrete 
within the framework of the efficient Hill accretion. The accretion rate is additionally a function of the pebble scale height given by
\begin{equation}
\label{eq:Hpebble}
 H_{\rm peb} = H_{\rm g} \sqrt{\frac{\alpha}{{\mathcal S}}} \ ,
\end{equation}
where $\alpha$ is the viscosity parameter and $\mathcal S$ the Stokes number. In case the Hill radius of the planet is larger than the 
pebble scale height, the accretion rate of the planet is given by \citep{LJ14}
\begin{equation}
\label{eq:Mdotpebble}
 \dot{M}_{\rm c, 2D} = 2 \left(\frac{{\mathcal S}}{0.1}\right)^{2/3} r_{\rm H} v_{\rm H} \Sigma_{\rm peb} \ ,
\end{equation}
where $r_{\rm H} = r [M_{\rm c} / (3M_\star)]^{1/3}$ is the Hill radius, $v_{\rm H}=\Omega_{\rm K} r_{\rm H}$  the Hill speed, and 
$\Sigma_{\rm peb}$  the pebble surface density. If the Stokes number of the particles exceeds 0.1, the accretion rate is 
limited to
\begin{equation}
 \dot{M}_{\rm c, 2D} = 2 r_{\rm H} v_{\rm H} \Sigma_{\rm peb} \ ,
\end{equation}
because the planetary seed cannot accrete particles from outside its Hill radius \citep{LJ12}. However, if the planetary seeds are 
small (as in the beginning of growth), the planet's Hill radius is smaller than the pebble scale height of the disc, meaning that 
the planet is fully embedded in the flow of pebbles and the accretion rate changes to \citep{Morby15}
\begin{equation}
 \dot{M}_{\rm c, 3D} = \dot{M}_{\rm c, 2D} \left( \frac{\pi ({\mathcal S}/0.1)^{1/3} r_{\rm H}}{2 \sqrt{2 \pi} H_{\rm peb}} \right). 
\end{equation}
The transition from 3D to 2D pebble accretion is then reached when
\begin{equation}
 \label{eq:2D3D}
 \frac{\pi ({\mathcal S}/0.1)^{1/3} r_{\rm H}}{2 \sqrt{2 \pi}} > H_{\rm peb} \ .
\end{equation}
This transition depends on the scale height of the disc and on the particle size, indicating that a higher planetary mass is needed to 
reach the faster 2D accretion branch in the outer parts of the disc where $H_{\rm peb}$ is larger. The Stokes number of the dominant 
particle size given by
\begin{equation}
 {\mathcal S} = \frac{\sqrt{3}}{8} \frac{\epsilon_{\rm P}}{\eta} \frac{\Sigma_{\rm peb}}{\Sigma_{\rm g}} \ .
\end{equation}
Here we set the parameters $\epsilon_{\rm P}$ to  $0.5$ and $\epsilon_{\rm D}$ to $0.05$ as in \citet{LJ14}. The final Stokes number 
obtained is constrained by the equilibrium between growth and drift to fit advanced coagulation models \citep{Birnstiel12}. The 
pebble surface density depends on the gas surface density $\Sigma_{\rm g}$ and the semi-major axis $r_{\rm p}$ through
\begin{equation}
 \label{eq:SigmaPeb}
 \Sigma_{\rm peb} = \sqrt{\frac{2 \dot{M}_{\rm peb} \Sigma_{\rm g} }{\sqrt{3} \pi \epsilon_{\rm P} r_{\rm P} v_{\rm K}}} \ ,
\end{equation}
where the pebble flux is
\begin{equation}
 \dot{M}_{\rm peb} = 2 \pi r_{\rm g} \frac{dr_{\rm g}}{dt} (Z_{\rm peb} \Sigma_{\rm g}) \ .
\end{equation}
Here $Z_{\rm peb}$ denotes the fraction of solids (metallicity) in the disc that can be transformed into pebbles at the pebble 
production line $r_{\rm g}$ at time $t$
\begin{equation}
 r_{\rm g} = \left(\frac{3}{16}\right)^{1/3} (GM_\star)^{1/3} (\epsilon_{\rm D} Z)^{2/3} t^{2/3} \ ,
\end{equation}
and
\begin{equation}
 \frac{dr_{\rm g}}{dt} = \frac{2}{3} \left(\frac{3}{16}\right)^{1/3} (GM_\star)^{1/3} (\epsilon_{\rm D} Z)^{2/3} t^{-1/3} \ ,
\end{equation}
where $M_\star$ is the stellar mass, which we set to $1 M_\odot$. \\

This pebble growth mechanism in combination with planet migration and disc structure evolution was used in \citet{Bitsch15} to 
constrain the growth of planets in discs, and the same code is used here as well. We just adapt it to the different disc models and 
viscosities, but the pebble accretion model stays the same.

\end{document}